\documentclass[%
 reprint,
 amsmath,amssymb,
 aps,
]{revtex4-1}

\usepackage{graphicx}
\usepackage{dcolumn}
\usepackage{bm}
\usepackage{hyperref}

\begin{document}

\title{Stochastic model of randomly end-linked polymer network micro-regions. }

\author{Sam C.P. Norris} 
\author{Andrea M. Kasko} 
\affiliation{Department of Bioengineering, University of California Los Angeles,Los Angeles, California, USA} 
\author{Tom Chou} 
\affiliation{Department of Computational Medicine}
\affiliation{Department of Mathematics, University of California Los Angeles, Los Angeles, California, USA}

\author{Maria R. D'Orsogna} 
\email{dorsogna@csun.edu}
\affiliation{Department of Mathematics, California State University Northridge, Northridge, California, USA} 

\date{\today}

\begin{abstract}
Polymerization and formation of crosslinked polymer networks are
important processes in manufacturing, materials fabrication, and in
the case of hydrated polymer networks, synthesis of biomedical
materials, drug delivery, and tissue engineering.
While considerable research has been devoted to the modeling of
polymer networks to determine averaged, mean-field, global properties,
there are fewer studies that specifically examine the variance of the
composition across ``micro-regions'' (composed of a large, but finite,
number of polymer network strands) within the larger polymer network.
Here, we mathematically model the stochastic formation of polymer
networks comprised of linear homobifunctional network strands that
undergo an end-linking gelation process.  We introduce a master
equation that describes the evolution of the probabilities of possible
network micro-region configurations as a function of time and extent
of reaction. We specifically focus on the dynamics of network
formation and the statistical variability of the gel micro-regions,
particularly at intermediate extents of reaction. We also consider
possible annealing effects and study how cooperative binding between
the two end-groups on a single network-strand affects network
formation.
Our results allow for a more detailed and thorough understanding of
polymer network dynamics and variability of network properties.
\end{abstract}

\maketitle

\section{Introduction}
The study of crosslinked polymer networks is important in many
applications from heavy industry to biomedical research
\cite{DeGennes1979,Flory1953,rubinstein2003,Peppas2006,Stauffer1982,
  Sangeetha2005,Laftah2011,Nicolson2001}. Crosslinked polymer networks
can be formed by various techniques, leading to a diverse and complex
set of structures and properties. Of these network types, considerable
attention has been paid to those formed by a process termed
``end-linking''. End-linked networks are usually comprised of
polymeric precursors, or ``network strands,'' that contain $N$
reactive end-groups\cite{Hild1998,Mark1977}. During
gelation, crosslinks, or ``branchpoints,'' link multiple end-groups
together. For example, poly(ethylene glycol) (PEG)-based hydrogels,
which are common in biomedical applications, are typically formed
through the reaction of its end-groups~\cite{Lin2009a}.

Network strands can bind to form a network via two main polymerization
reaction mechanisms: step-growth and chain-growth.  Several excellent
reviews have been written to describe these
processes~\cite{Lin2009a,Labana1984,Mespouille2009,Buwalda2014}.
Briefly, gelation by step-growth polymerization
  typically involves a defined binary reaction (e.g., thiol-ene or
  azide-alkyne reactions) between the network strand end-groups and
  the complementary binding sites of a multifunctional branchpoint
  which acts to crosslink the network strands together. Networks
  formed by step-growth polymerization are typically more homogeneous
  in structure since the number of functional groups per branchpoint
  can be fixed.  Gelation by chain-growth polymerization occurs via a
  chain-extension reaction where the network strand end-groups bind to
  a growing chain of end-groups, termed the ``active center'' (e.g.,
  free-radical polymerization of vinyl end-groups). The chain of
  end-groups forms a branchpoint that crosslinks the network strands
  together. Networks formed by chain-growth polymerization tend to
  have a more heterogeneous structure since the number of end-groups
  bound at the branchpoint is not fixed.

\begin{figure*}[t!]
\centering
\includegraphics[width=\textwidth]{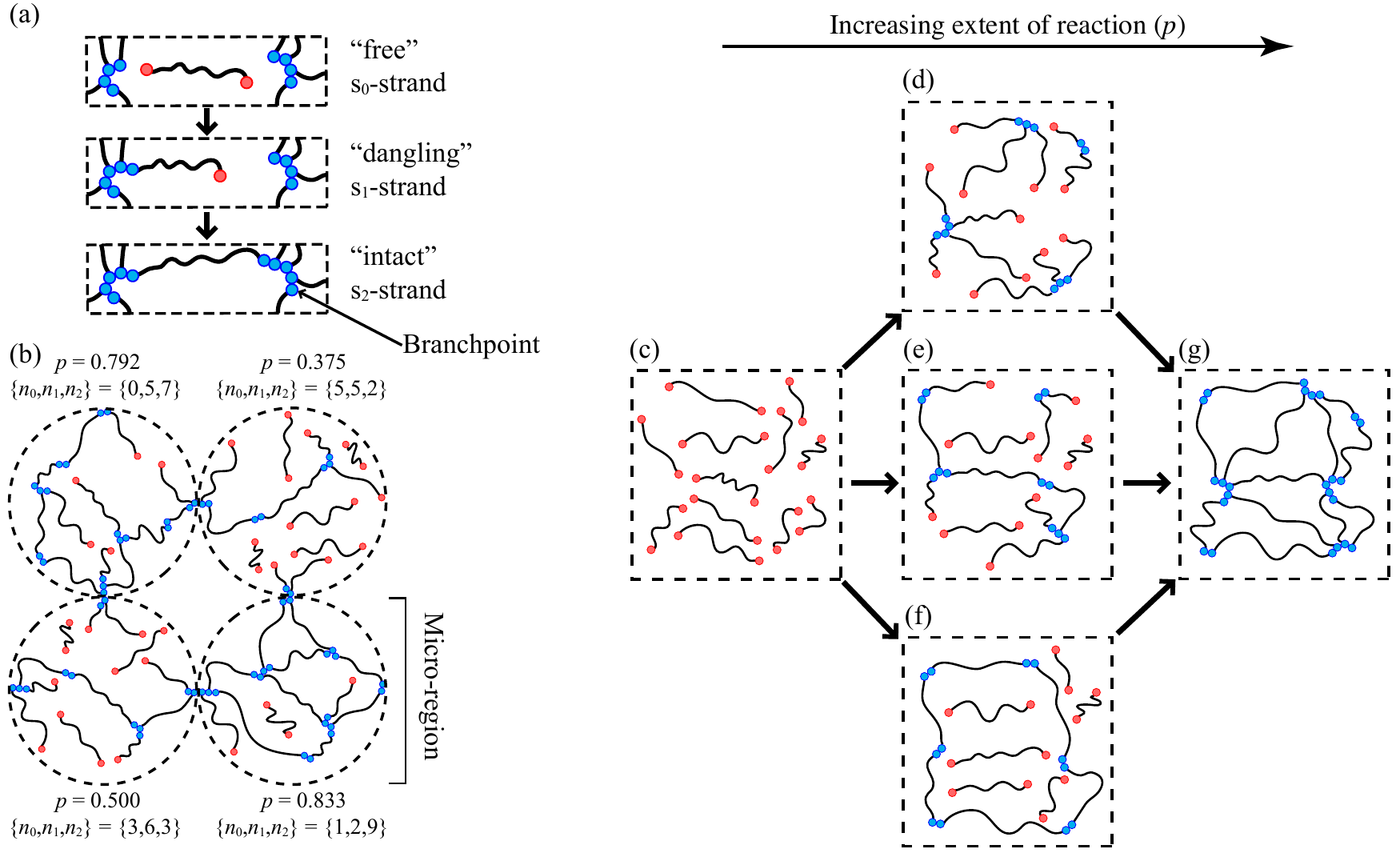}
      \caption{Polymer network configurations. Black lines represent
        network strands with $N=2$ reactive end-groups that can be
        unbound (red dots) or bound (blue dots). Reactive ends bind to
        form a branchpoint connecting different strands.  \textbf{(a)}
        Bifunctional strands are either ``free'' -- neither end-group
        is bound ($s_\text{0}\text{-strand}$); ``dangling'' -- a
        single end-group is bound ($s_\text{1}\text{-strand}$); or
        ``intact'' -- both end-groups are bound
        ($s_\text{2}\text{-strand}$).  \textbf{(b)} Schematic of four
        uncoupled ``micro-regions'' (dashed lines) within the
        network. Each is comprised of $N_\text{s}$ strands but the
        extent of reaction $p$ can vary.  \textbf{(c)} For $p=0$ all
        network strands are in the
        $s_\text{0}\text{-state}$. \textbf{(d-f)} For intermediate
        $0<p<1$, many configurations are possible, including
        \textbf{(d)} only $s_\text{1}\text{-strands}$, \textbf{(e)} a
        combination of $s_\text{0}\text{-}$, $s_\text{1}\text{-}$, and
        $s_\text{2}\text{-}$strands, or \textbf{(f)} only
        $s_\text{0}\text{-}$ and
        $s_\text{2}\text{-}$strands. \textbf{(g)} For $p=1$, the only
        possible configuration is for strands be fully bound in the
        $s_\text{2}\text{-state}$.}
      \label{fgr:NetworkFormationDegradation}
\end{figure*}

Network strands binding via either step- or chain-growth may exist in
many states. For bifunctional strands with $N=2$ reaction sites three
possibilities arise as depicted in Figure~\ref{fgr:NetworkFormationDegradation}a: (\textit{i}) the strand may be
``free'' where neither of the reactive ends have bound ($s_0$-strand);
(\textit{ii}) the strand may ``dangle'' where only a single end has
bound and the strand dangles from the rest of the network
($s_1$-strand); or (\textit{iii}) the strand may be ``intact'' where
both ends are bound to the larger polymer network and bridge two
different branchpoints ($s_2$-strand) \cite{Metters2000a}.  Strands
with both ends bound may also form a loop, where both ends are bound
to the same branchpoint \cite{Tonelli1974}. The proportion of free,
dangling and intact network strands may affect the chemical and
physical properties of the network, for example in water-swollen
polymeric networks, bound strand ratios impact gel modulus, mesh
size, and swelling\cite{rubinstein2003}.

Finally, the architecture of polymer networks formed by end-linking
gelation is not spatially uniform. Heterogeneous domains within
polymer networks exist that span a few to hundreds of nanometers in
size and arise through variations in local strand concentration
(termed ``frozen concentration fluctuation''), heterogeneous
distribution of crosslinking, or topological- and connectivity-based
inhomogeneities due to variability in network strand assembly
\cite{Gu2019a, Seiffert2017, Bastide1988, Ikkai2005}.  We define these
microscopic domains as ``micro-regions.'' For simplicity we assume
micro-regions are statistically identical, independent, and
composed of a fixed number $N_{\rm s}$ of strands. Each strand is also
assumed to carry $N = 2$ reactive end groups, the most representative
experimental scenario \cite{Hild1998}, resulting in a total of $2
N_\text{s}$ available binding sites per micro-region.  We also denote
by $m$ the number of bound end-groups in each micro-region, so that the fraction
$p$ of bound end-groups per micro-region is $p = m/(2 N_\text{s})$.
This quantity is also known as the extent of reaction, and can be
experimentally tuned to control the elastic modulus, viscosity,
swelling, mesh size, and other network properties\cite{Flory1953}.
By definition $0 \leq p \leq 1$ since the number of bound end groups
$m$ cannot exceed the total number of available ones $2 N_\text{s}$.
Finally we assume that micro-regions are large enough that
  boundary effects between bordering domains are negligible so that
  the free, dangling, intact strand distribution of two adjacent
  micro-regions are not correlated. Note that the same value of $0 <
p < 1$, may be associated to different $\{n_0, n_1, n_2 \}$
micro-region configurations with $n_0$ free, $n_1$ dangling, and $n_2$
intact strands.  Figure~\ref{fgr:NetworkFormationDegradation}b shows
four different micro-region realizations within a larger network where
$N_{\rm s}$ is fixed but different $\{n_0, n_1, n_2 \}$ configurations
arise, resulting in different extents of reaction $p$.  In
Figure~\ref{fgr:NetworkFormationDegradation}d-f we show several
distinct $\{n_0, n_1, n_2 \}$ configurations corresponding to fixed
$N_{\text s}$ and $p$.

Some quantities of interest may be derived using $p$ such as the
likelihood $P_{\ell}\big(p \big)$ \cite{DeGennes1979, Flory1953,
  Bibbo1982, Dusek2002, Metters2000a} of finding free ($\ell=0$),
dangling $(\ell =1)$, and intact ($\ell=2$) strands for a given $p$.  A
stochastic analysis however would lead to an expression for the
probability distribution of finding any micro-region configuration
$\{n_0, n_1, n_2 \}$ corresponding to a given $p$, offering a much
richer understanding of the binding process.  Previously developed
stochastic models use a subset-of-states approach\cite{Stanford1975,
  Stanford1975a, Ahmad1980}, where a polymerizing mixture is described
as a set of ``subgraph'' states of monomeric strands, a subset of
which is used to drive polymerization\cite{Schamboeck2019,Wang2016a}.
These models, however, only examine the connectivity of small subgraphs,
typically made of only a few network strands, to represent large scale networks and
predict bulk quantities such as the network gel point \cite{Lin2018,
  Schamboeck2019}.  Studies involving larger subgraphs containing a
sizable number of strands (say, greater than ten)  are still
lacking. Finally, although several Monte Carlo numerical studies have
examined network heterogeneity\cite{Kroll2017, Kroll2015,
  Balabanyan2005, Gilra2000, Leung1984a, Leung1984, Hosono2007}, none
of them have evaluated configuration probability distributions.

We aim to determine the probability distribution for a given
configuration of free, dangling, and intact strands within a
micro-region of $N_{\text s}$ bifunctional strands, given a total
number of $m$ bound end-groups, or equivalently for fixed $p = m /2
N_{\text s}$.  This will allow us to go beyond the characterization of
a micro-region by means of $p$ and $N_{\text s}$ alone and to obtain
analytical expressions for micro-region properties that depend on
possible $\{n_0, n_1, n_2 \}$ configurations.  In some experimental
scenarios $m$ or $p$ may change among micro-region realizations, and
it may be useful to understand the structure of the network for a
given average, intermediate extent of reaction $\langle p \rangle$.
Our results will thus be presented both as a function of time, and of
the average extent of reaction. We draw on existing stochastic self
assembly and nucleation models\cite{Chou2011, DOrsogna2011,
  Yvinec2012, DOrsogna2013, Davis2016} and utilize a master equation approach.
Different forms of the master equation will be developed and analyzed
to account for different end-group reactivities and the possibility
for end-groups to dynamically rearrange within the micro-region.  We
do not model branchpoint functionality but focus on the number of
intact, dangling, and free $\{n_0, n_1, n_2 \}$ strands within
micro-regions.  As a result, the total number of branchpoints and
topology of the network do not affect our modeling, so while the
structures depicted in Figure~\ref{fgr:NetworkFormationDegradation}
resemble those formed by chain-growth polymerization, our methods can
be easily applied to step-growth polymerization as well.
Table~\ref{tbl:Variables_Summary} lists the various quantities used in
the remainder of this work.
\begin{table*}
\caption{Summary of variables
  used.}\label{tbl:Variables_Summary}
\begin{tabular*}{0.6\textwidth}{@{\extracolsep{\fill}}cl}
\hline \hline
Symbol  & Representation        \\ \hline 
$N$ & Number of reactive end-groups per network strand\\
$\ell$ & Number of bound end-groups  per network strand\\
$s_\ell$ & Designation of strand type with $\ell$ bound end-groups  \\
$N_\text{s}$ & Number network strands per micro-region\\ 
$n_0$ & Number of $s_\text{0}\text{-network}$ strands per micro-region \\ 
$n_\text{1}$ & Number of $s_\text{1}\text{-network}$ strands per micro-region \\ 
$n_2$ & Number of $s_\text{2}\text{-network}$ strands per micro-region \\ 
$m$ & Total number of bound end-groups  per micro-region\\ 
$p$ & Extent of reaction, $m/(N_\text{s} N)$\\ 
$t$ & Time of reaction (time units)\\
$P(n_1,n_2,t)$ & Micro-region  configuration  $\{n_0,n_1,n_2\}$ probability at  $t$   \\
$\alpha$ &  Reactivity/cooperative binding parameter (unitless)\\
$\lambda$ & Binding rate ($\text{time}^{-1}$)\\
$\kappa$ & End-group rearrangement rate to binding rate ratio (unitless)\\
\hline \hline
  \end{tabular*}
\end{table*}

\section{Mathematical Models and Analysis}

For completeness, we first review basic combinatoric and equilibrium
models describing network formation before introducing our master
equation models.

\subsection{Combinatoric and equilibium models}
\label{Sec:Combinatoric}

Many mathematical studies of network formation via end-linking have
used combinatoric approaches to quantify the number of polymeric
strands in a given state
\cite{Miller1976,Macosko1976a,Metters2000a,Tibbitt2013b,Norris2017}.
The extent of reaction $p$, defined as the fraction of bound end-groups
per micro-region, can also be interpreted as the probability that any
end-group within a micro-region has bound. The probability
$P_{\ell}(p)$ of finding an $s_\ell$-strand with $0 \leq
\ell \leq N $ bound end-groups is thus
\begin{equation}
\label{eqn:binomial}
P_{\ell}(p) = {N \choose \ell} {p}^\ell \big (1-p \big)^{N-\ell},
\end{equation}
which assumes that of $N$ end-groups, $\ell$ are bound and $N-\ell$
are not.  Equation~\ref{eqn:binomial} provides a basis for mean field
end-linking gelation models used to predict network
properties. Henceforth we assume $N=2$.  The average number
$n_{\ell}(p)$ of $s_\ell$-strands within a micro-region of $N_{\rm s}$
strands can thus be written as
\begin{equation}
\label{eqn:avg}
n_{\ell}(p) = N_{\text s} {2 \choose \ell} {p}^\ell (1-p)^{2-\ell}.
\end{equation}
Equation~\ref{eqn:avg} does not provide any information on the
possible spatial arrangement of free, dangling, intact, strands within a
micro-region. Some end-groups may also bind differently than others
depending on their state.  For example free strands might more readily
bind than dangling ones since diffusion allows them to more easily
navigate the local environment to find an appropriate reaction
site. Cooperative binding arises when the unbound end-group of a
dangling strand more readily binds to form a fully bound, intact
strand due to its proximity to the polymerizing network, especially
when the polymer solution is dilute.  Uncooperative binding emerges
when $s_\text{2}\text{-strand}$ formation from the binding of an
existing $s_\text{1}\text{-strand}$ is hindered by negative allosteric
effects, which has been shown to occur in rigid strands
\cite{Hosono2007}. Finally, the reaction steps associated with network
formation may also be irreversible or reversible. Irreversible
reactions lead to ``quenched'' network formation whose properties are
highly dependent on initial conditions, while reversible reactions
allow the network to rearrange while forming and ``anneal.''


The binding scenarios described above lead to different probability
distributions for a given micro-region configuration.  Evaluating
these distributions requires a more complex mathematical
representation than Equations~\ref{eqn:binomial} or \ref{eqn:avg}.
Some can be can be derived via combinatoric arguments, for example in
the case of reversible binding, when equilibrium is reached and the
annealing process is complete. We evaluate such limit here and find
the probability distribution for a given micro-region configuration
$\{n_0, n_1, n_2 \}$ with $n_{0}$ unbound $s_\text{0}\text{-strands}$,
$n_{1}$ singly bound $s_\text{1}\text{-strands}$, and $n_{2}$ doubly
bound $s_\text{2}\text{-strands}$, under the assumption that a total
of $m$ end-groups have bound.

\begin{figure*}[t!]
\centering
\includegraphics[width=6in]{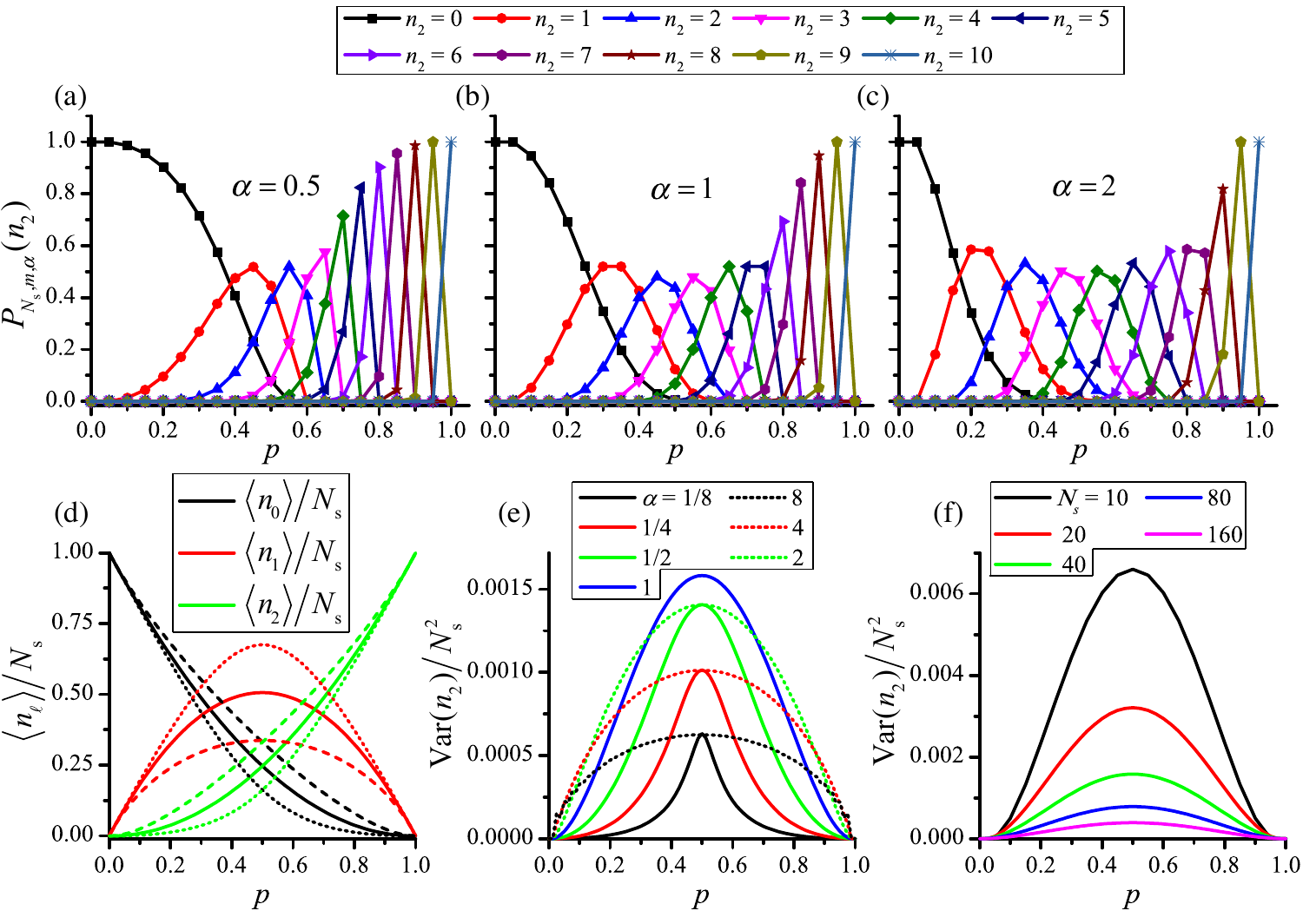}
\caption{Results from the equilibrated distribution $P_{N_\text{s},m,
    \alpha}(n_2)$ as evaluated from Equation~\ref{eqn:pbias} for
  $N_{\rm s} = 10$ and as a function of the extent of reaction $p =
  m/2 N_{\rm s}$ for \textbf{(a)} $\alpha=0.5$, \textbf{(b)}
  $\alpha=1$, and \textbf{(c)} $\alpha=2$. Each point represents a
  different configuration $\{n_0,n_1,n_2\}$; those with the same $n_2$
  are connected by lines.  \textbf{(d)} Average populations $\langle
  n_{\ell}\rangle/N_\text{s}$ for $\ell = 0,1,2$, $N_\text{s}=40$ and
  $\alpha=0.5$ (dotted line), $\alpha=1$ (solid line) and $\alpha=2$
  (dashed line).  \textbf{(e)} The normalized variance of $n_2/N_{\rm
    s}$ as a function of $p$ for $N_\text{s}=40$, and several values
  of $\alpha$.  \textbf{(f)} The normalized variance of $n_2/N_{\rm
    s}$ for $\alpha =1$, and several values of $N_{\rm s}$.  A maximum
  emerges at $p = 0.5$, whose value decreases with $N_{\rm s}$.
  Notice the symmetry in $\langle n_1 \rangle$ and
  $\mathrm{Var}(n_2)/N_{\rm s}^2$ about $p=0.5$ for all values of
  $\alpha$.}
\label{fgr:PartFunc}
\end{figure*}

At equilibrium, the time and order at which strands were linked do not
affect configuration likelihoods, so the task of finding the
probability distribution for $\{n_0, n_1, n_2 \}$ is equivalent to
finding the number of ways ${\cal N}(n_0, n_1, n_2)$ one can
distribute $\{n_0, n_1, n_2 \}$ among $N_\text{s}$ strands with $m$
total bindings. The above quantities are related by $n_0+ n_1 + n_2 =
N_\text{s}$ since all strands must be accounted for, and by $n_1 + 2
n_2 = m$ to include the contribution of each strand type to the total
bound end-group count.  Hence, a given micro-region with configuration
$\{n_0, n_1, n_2 \}$ can be equivalently described by $\{N_\text{s},
m, n_2 \}$.  The extent of reaction $p$ can also be determined from
$\{N_\text{s}, m, n_2 \}$ via $p = m/(2 N_\text{s} )$.  Combinatoric
arguments yield ${\cal N}(n_0, n_1, n_2)$ as
\begin{equation}
{\cal N}(n_0, n_1, n_2)= 2^{n_1} {N_\text{s} \choose n_0 \, n_1 \,n_2}.
\end{equation}
\noindent
Here, the $2^{n_1}$ factor arises from the fact that the bound end-group
on an $s_\text{1}\text{-strand}$ can be arranged in two 
configurations per strand.  The above can be rewritten 
using $n_0 =N_\text{s}-m+n_2 $ and $n_1 = m-2n_2$ as follows
\begin{eqnarray}
{\cal N}(N_\text{s}, m, n_2)= 
 \frac {2^{m-2n_2}N_\text{s}!}{(N_\text{s}-m+n_2)!(m-2n_2)! n_2! }.
\label{eqn:PartialProb}
\end{eqnarray}
\noindent
Upon summing over $n_2$ with  $N_\text{s}, m$ fixed, we find $Z_{N_\text{s},m}$ the partition function
over all possible
configurations, with $N_\text{s}, m$ fixed
\begin{equation}
Z_{N_\text{s},m} =   \sum_{n_2=0}^{[m/2]} {\cal N}(N_\text{s}, m, n_2),
\end{equation}

\noindent
where $[\cdot]$ indicates the integer part of its argument.  The equilibrium
probability distribution can finally be calculated as
\begin{eqnarray}
\label{eqn:pcomb}
P_{N_\text{s},m}(n_2) = \frac{ {\cal N}(N_\text{s}, m, n_2)} {Z_{N_\text{s}, m}}.
\end{eqnarray}
Equation~\ref{eqn:pcomb} may be used to evaluate many different
micro-region properties, such as averages, variances, and higher
moments.  We begin with the average number of free, dangling,
and intact strands, respectively given by

\begin{subequations}
\label{eqn:AveragePartFunc}
\begin{align}
\langle n_{0}\rangle &=N_\text{s}-m+\sum_{n_2=0} ^{[m/2]} n_2 P_{N_\text{s}m} (n_2).\\
\langle n_{1}\rangle &=m-2 \sum_{n_2=0} ^{[m/2]} n_2 P_{N_\text{s}, m} (n_2),\\
\langle n_{2}\rangle &=\sum_{n_2=0} ^{[m/2]} n_2 P_{N_\text{s}, m} (n_2).
\end{align}
\end{subequations}
In Equations~\ref{eqn:AveragePartFunc} the average, denoted by
$\langle \cdot \rangle$, is taken across all micro-regions with the
same $N_s$ and $m$, or equivalently, using all possible
  configurations within a single micro-region with $N_{\text s}$
  strands and $m$ total number of bound end-groups.
The above combinatoric argument assumes that
end-group binding is accompanied by end-group annealing until
equilibrium is reached, independent of the number of bound end-groups
already present.  However, within cooperative or uncooperative
binding, bound end-groups may promote or hinder the binding of other
end-groups.  We include these phenomena by re-writing
Equation~\ref{eqn:PartialProb} as
\begin{equation}
\label{eqn:bias}
{\cal N}(N_\text{s}, m, n_2, \alpha) =  \frac {(2/\alpha)^{m-2n_2}N_\text{s}!}{(N_\text{s}-m+n_2)!(m-2n_2)!
n_2! },
\end{equation}
where the reactivity parameter $\alpha > 1$ represents cooperative binding, penalizing
dangling ends in favor of $s_\text{1}\rightarrow s_\text{2}$ events.  Values of $\alpha<1$ represent
uncooperative binding where $s_\text{0}\rightarrow s_\text{1}$
events are favored.  The neutral case is $\alpha =1$. Finally, the 
equilibrium probability distribution  $P_{N_\text{s},m,\alpha}(n_2)$ can be written as
\begin{equation}
\label{eqn:pbias}
P_{N_\text{s},m,\alpha}(n_2) = \frac{{\cal N}({N_\text{s}},m,{n_2},\alpha)}{\sum\limits_{{n_2}
= 0}^{[m/2]} {\cal N} ({N_\text{s}},m,{n_2},\alpha)}. 
\end{equation}
We plot $P_{N_\text{s},m, \alpha}(n_2)$ in Equation~\ref{eqn:pbias}
for several values of $n_2$, under three choices of $\alpha $ and as a
function of the extent of reaction $p= m/(2 N_\text{s})$ in
Figures~\ref{fgr:PartFunc}a--c.  The solid lines connect micro-region
configurations with the same $n_2$; we choose this representation as
the number of intact ``elastically effective''
$s_\text{2}\text{-network}$ strands is an important feature of polymer
networks and determines both the mechanical modulus and swelling
behavior of the network \cite{rubinstein2003}.
As $\alpha$ increases, all curves tend to shift to the left, as might
be expected since increasing cooperative effects favor the emergence
of $s_2$-strands for a given $p$.  In Figure~\ref{fgr:PartFunc}d we
plot the average strand fractions $\langle n_{\ell}\rangle/N_\text{s}$
for $\ell = 0,1,2$ as evaluated via Equations~\ref{eqn:AveragePartFunc} for $N_\text{s}=40$ and as a function of
$p$.  Note that for any $\alpha$, the average quantity $\langle
n_{1}\rangle$ is a symmetric function of $m$ about $N_{\rm s}$ as can
be verified by imposing $m' = 2N_{\rm s} - m$ in
Equation~\ref{eqn:AveragePartFunc}b and verifying that $\langle
n_{1}\rangle$ remains unchanged. Since $p = m /2 N_{\rm s}$, this also
implies that $\langle n_{1}\rangle$ will be symmetric about $p=1/2$
for all values of $\alpha$, as seen in Figures~\ref{fgr:PartFunc}e-f.
We also calculate the second moment $\langle n_2^2\rangle$ defined as
\begin{equation}
\label{eqn:SecondMomentPF}
\langle n_2^2\rangle = \sum_{n_2=0}^{[m/2]}  n_2^2
P_{N_\text{s},m,\alpha}(n_2), 
\end{equation}
from which we obtain the variance $\mathrm{Var}(n_2) = \langle n_2^2
\rangle - \langle n_2\rangle^2$ where $\langle n_2\rangle$ is derived
in Equation~\ref{eqn:AveragePartFunc}c.  Similarly as for $\langle n_1
\rangle$ one can verify that $\mathrm{Var}(n_2)$ is symmetric about
$p=1/2$ for all values of $\alpha$.  Since $n_1 = m-2n_2$, $n_0 =
N_\text{s}-m+n_2$, and given $\langle n_1\rangle$ and $\langle
n_0\rangle$ from Equations~\ref{eqn:AveragePartFunc}a-b,
$\mathrm{Var}(n_1) = \langle n_1^2 \rangle - \langle n_1\rangle^2$ and
$\mathrm{Var}(n_0)= \langle n_0^2 \rangle - \langle n_0\rangle^2$ can
also be derived using Equations~\ref{eqn:AveragePartFunc}c and
\ref{eqn:SecondMomentPF}.  Figure~\ref{fgr:PartFunc}e shows
$\mathrm{Var} (n_{2})/N_{\rm s}^2$ as a function of $p$ for different
values of $\alpha$.  In each case, the maximum variance occurs when
half of all possible end-groups have bound at $p=1/2$.  As $\alpha$
deviates from the neutral condition $\alpha =1$, the bias towards
certain bond types induced by cooperativity or uncooperativity causes
the variance to decrease.  In Figure~\ref{fgr:PartFunc}f we plot
$\mathrm{Var}(n_2)/N^2_\text{s}$ as a function of $p$ for different
values of $N_\text{s}$: the curve remains symmetric about $p=0.5$ and
as $N_s$ increases, the normalized variance decreases.
\subsection{Dynamic models: Master Equation approaches}

We now derive the probability distribution $P(n_0, n_1, n_2,t)$ of
finding a given $\{n_0, n_1, n_2,t\}$ micro-region configuration at
time $t$ through a master equation that allows for the inclusion of
reversible/irreversible (annealed/quenched) bond formation, and
cooperative/uncooperative binding.  Since the total number of strands
per micro-region is constant, the constraint $n_0 + n_1 + n_2 =
N_\text{s}$ is obeyed at all times and effectively $P(n_0, n_1, n_2,
t) \rightarrow P(n_1, n_2, t)$.  We compare equilibrium or steady
state solutions to Equation~\ref{eqn:binomial}; where possible we
also determine the full time-dependent solution for $P(n_1, n_2, t)$,
which can be used to derive other quantities of interest, such as the
variance and higher moments.
\subsubsection{Quenched end-group binding}
\label{quenched}
\begin{figure}[t!]
\centering
\includegraphics[width=3.375 in]{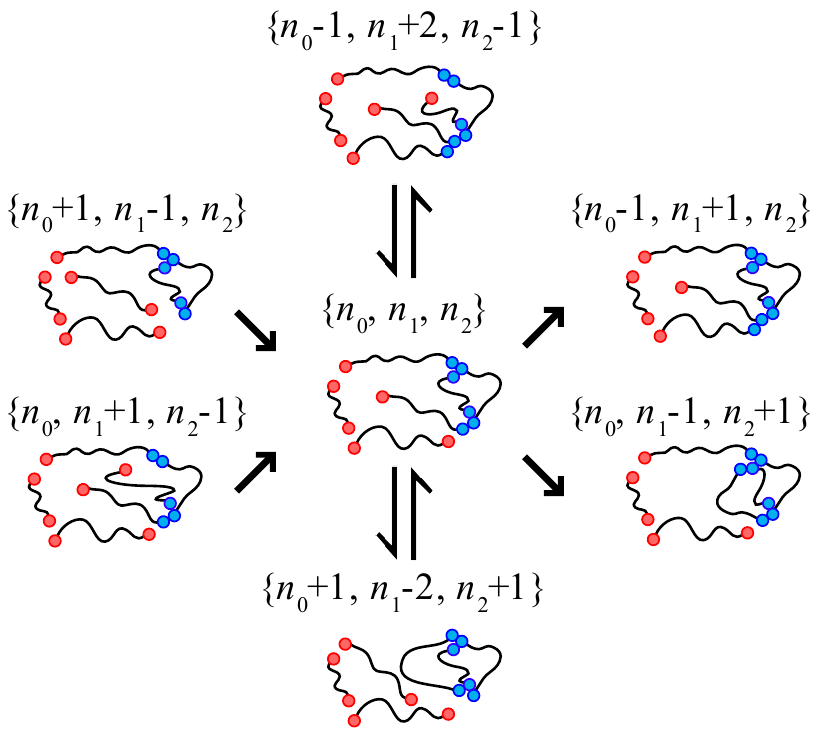}
      \caption{Possible end-group binding transitions for $N_\text{s}
        =6$. The state at the center of the schematic is
        $\{n_0,n_1,n_2\}=\{2,2,2\}$ corresponding to $m=6$. To the
        left are two $m=5$ and to the right are two $m=7$
        configurations. Under quenched binding discussed in
        Section~\ref{quenched} the dynamics will flow from left to
        right following the unidirectional arrows. To the top and
        bottom of the $\{n_0,n_1,n_2\}=\{2,2,2\}$ state are other
        $m=6$ configurations. Under dynamic rearrangement discussed in
        Section~\ref{dynamic} the system equilibrates following the
        vertical lines.}
      \label{fgr:s0-s1-s2_transition}
\end{figure}

\begin{figure}[t!]
\centering
\includegraphics[width=3.375 in]{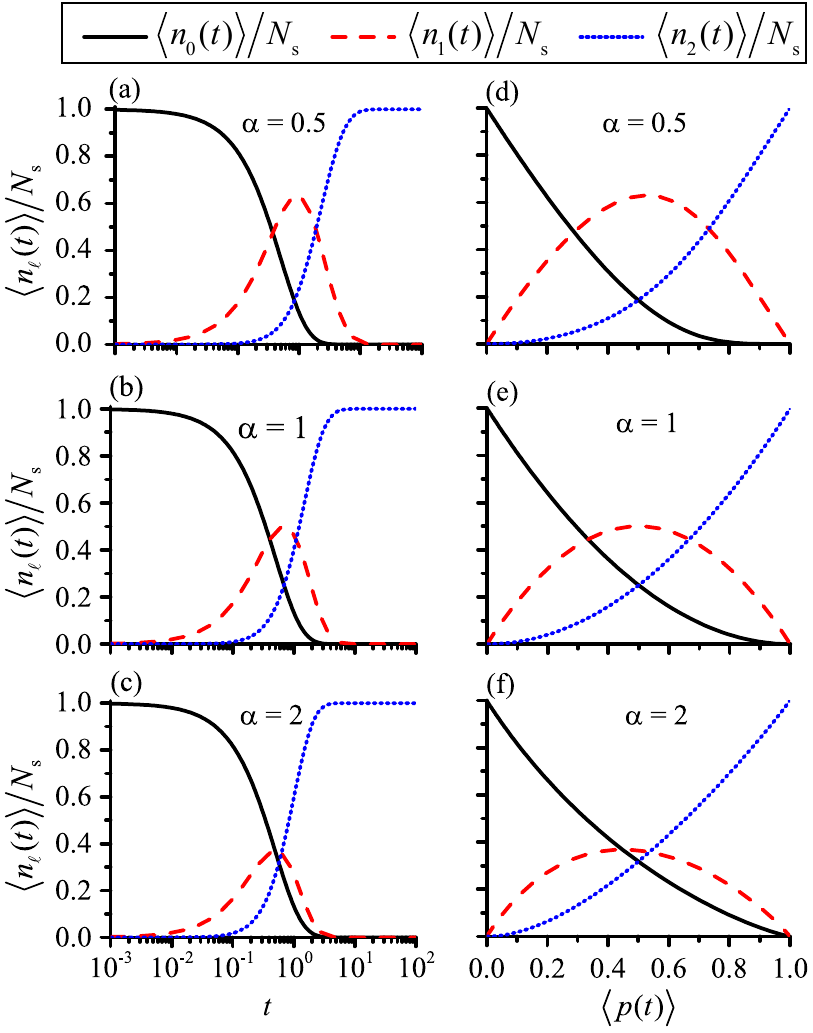}
\caption{Average strand fractions $\langle n_\ell(t) \rangle
  /N_\text{s}$ for $\ell = 0,1,2$ and $N_{\rm s} =40 $ as evaluated 
  from Equations~\ref{eqn:MassActionSol} and plotted as a function of
  \textbf{(a-c)} time and \textbf{(d-f)} parametrically against the
  extent of reaction $\langle p(t) \rangle$ given by
  Equation~\ref{eqn:poft}.  The chosen values of the reactivity
  parameter $\alpha$ are: \textbf{(a,d)} $\alpha=0.5$, \textbf{(b,e)}
  $\alpha=1$, and \textbf{(c,f)} $\alpha=2$.}
\label{fgr:Meq}
\end{figure}

The first case we consider is that of irreversible (or quenched)
end-group binding, whereby once an end-group has bound, it will not
detach. We also assume the binding rate $\lambda$ of an end-group is
constant.  Under these conditions, the master equation for the
probability distribution $P(n_1, n_2, t)$ evolves according to
\begin{equation}
\begin{split}
\frac{\text{d}P(n_1, n_2, t)}{\text{d}t} = & 2 \lambda (N_\text{s} - n_1 - n_2 +1) P(n_1 - 1, n_2,t) \\
& + \lambda \alpha (n_1 +1) P(n_1 +1, n_2 - 1,t)  \\ 
& - \lambda [2(N_\text{s}  - n_1 - n_2) + \alpha  n_1 ]  P(n_1, n_2,t), 
\end{split}
\label{eqn:MasterEqQuench}
\end{equation}
where we have explicitly used the $n_0 = N_\text{s} - n_1 - n_2$
constraint. Equation~\ref{eqn:MasterEqQuench} also includes the
reactivity parameter $\alpha$: $\alpha > 1$ represents cooperative
binding so that $s_\text{1} \rightarrow s_\text{2}$ binding events are
more likely than $s_\text{0} \rightarrow s_\text{1}$ events; the
reverse is true for uncooperative binding, $\alpha < 1$, where
$s_\text{0}\rightarrow s_\text{1}$ events are favored. The first term
on the right hand side of Equation~\ref{eqn:MasterEqQuench} represents
the process of an unbound strand attaching to the network structure to
form a singly bound dangling strand ($s_\text{0} \rightarrow
s_\text{1}$), which gives the configuration transition $\{n_0 + 1,
n_1-1, n_2\} \to \{n_0, n_1, n_2 \}$ (Figure~\ref{fgr:s0-s1-s2_transition}).
The multiplicative factor $N_\text{s} - n_1 - n_2 +1$ represents the
number $s_\text{0}\text{-strands}$ in the starting configuration that
can bind to the network; the two prefactor is included since an
$s_\text{0}\text{-strand}$ can bind to the network at either of its
two unbound end-groups. Similarly, the second term represents an
unbound end-group from a singly bound strand binding to the network
and forming a doubly bound strand ($s_\text{1} \rightarrow
s_\text{2}$). The related transition is $\{ n_0, n_1 +1, n_2-1 \} \to
\{n_0, n_1, n_2\}$ (Figure~\ref{fgr:s0-s1-s2_transition}). The
multiplicative factor $n_1+1$ represents the number of
$s_\text{1}\text{-strands}$ that can bind to the network to form an
$s_\text{2}\text{-strand}$.  Finally the last term describes the
processes that drives the system out of the $\{n_0, n_1, n_2 \}$
configuration, where either an $s_\text{0} \rightarrow s_\text{1} $
transition, with $\{n_0, n_1, n_2\} \to \{n_0-1, n_1+1, n_2 \}$, or an
$s_\text{1} \rightarrow s_\text{2} $ transition, with $\{n_0, n_1,
n_2\} \to \{n_0, n_1-1, n_2+1 \}$ occur
(Figure~\ref{fgr:s0-s1-s2_transition}).  The total number of distinct
$\{n_0, n_1, n_2 \}$ states can be enumerated via
\begin{eqnarray}
\label{eqn:count}
\sum_{n_2 = 0}^{N_{\rm s}} \sum_{n_1 = 0}^{N_{\rm s} - n_2} 1 = 
\frac{(N_{\rm s} + 2) (N_{\rm s} + 1)}2.
\end{eqnarray}
Due to the irreversibility of the binding process, at $t \to \infty$
we expect the system to consist only of $s_\text{2}\text{-network}$
strands: $P(n_1, n_2, t \to \infty) = 0$ for all $\{n_1, n_2\} \neq
\{0, N_\text{s}\}$ and $P(0, N_\text{s}, t \to \infty) = 1$ as
depicted in Figure~\ref{fgr:NetworkFormationDegradation}g.  We can
obtain an alternate representation for
Equation~\ref{eqn:MasterEqQuench} by using the $n_0+n_1+n_2=N_{\rm s}$
constraint to represent $n_2$ so that $P(n_0,n_1,n_2,t) \to
P(n_0,n_1,t)$ and the master equation reads
 \begin{equation}
 \begin{split}
\frac{\text{d}P(n_0, n_1, t)}{\text{d}t} = & 2 \lambda (n_0 +1)P(n_0 + 1, n_1-1,t)  \\
&  +  \lambda \alpha (n_1 +1) P(n_0, n_1 +1,t) \\ 
& - \lambda \left(2 n_0 + \alpha  n_1  \right) P(n_0, n_1,t).
\end{split}
\label{eqn:MasterEqQuenchedHybrid2}
\end{equation}
This representation is equivalent to
Equation~\ref{eqn:MasterEqQuench} and will be useful in deriving the
distribution $P(n_0, n_1,t)$ from which $P(n_1, n_2,t)$ can be
obtained. We now nondimensionalize our model by measuring time
  in units of the typical bond formation time, $\lambda^{-1}$.
  Henceforth, time $t$ will be dimensionless and $\lambda$ will no
  longer appear (equivalently, we set $\lambda = 1$ in
  Equations~\ref{eqn:MasterEqQuench} and
  \ref{eqn:MasterEqQuenchedHybrid2}). The mean number of strand types
  $\langle n_\ell (t) \rangle$ in a single micro-regions are defined
  by
\begin{equation}
\label{eqn:MassAction}
\langle n_\ell (t) \rangle =  \sum_{n_1, n_2} n_\ell P(n_1, n_2, t),
\end{equation}
for $\ell=1,2$, under the $0 \leq n_1 + n_2 \leq N_\text{s}$
constraint.  The corresponding mass-action equations can be derived by
multiplying Equation~\ref{eqn:MasterEqQuench} by $n_\ell$ and by
summing over $n_1, n_2$ under the same constraint so that
\begin{subequations}
\begin{align}
\frac{\text{d} \langle n_0 (t) \rangle}{\text{d}t} = & - 2\langle n_0 \rangle, \\
\frac{\text{d} \langle n_1 (t) \rangle}{\text{d}t} = & 2 \langle n_0  \rangle - {\alpha} \langle n_1 \rangle, \\
\frac{\text{d} \langle n_2 (t) \rangle}{\text{d}t} = &  \alpha \langle n_1 \rangle.
\end{align}
\label{eqn:MassAction_dt}
\end{subequations}
Equations~\ref{eqn:MassAction_dt}a-c can be solved under the initial
condition $n_0 (0) = N_\text{s}$, representing all strands being
unbound at $t=0$. We find
\begin{subequations}
\begin{align}
\langle n_0 (t) \rangle = & N_\text{s} e^{-2t }, \\
 \langle n_1 (t) \rangle = & N_\text{s} \frac{2\left(e^{- \alpha t}
-e^{-2t }\right) }{2-\alpha}, \\
  \langle n_2 (t) \rangle = &   N_\text{s} \left(1 + \frac{\alpha e^{-2t} - 2 e^{-\alpha t} }{2- \alpha} \right),
  \end{align}
  \label{eqn:MassActionSol}
\end{subequations}
\noindent
so that $\langle n_{\ell} (t \to \infty) \rangle \to 0$ for $\ell =
0,1$ and $\langle n_2 (t \to \infty) \rangle \to N_\text{s}$.
Equations~\ref{eqn:MassActionSol}a-c represent average values
calculated across all micro-regions at time $t$ under quenched
binding.  We compare the $t \to \infty$ limit of
Equations~\ref{eqn:MassActionSol}a-c to Equation~\ref{eqn:avg} which
estimates average strand numbers using combinatoric arguments.  To do
so, we evaluate $\langle m \rangle = \langle n_1 \rangle + 2 \langle
n_2 \rangle $ to find
\noindent
\begin{eqnarray}
\langle m (t) \rangle =  \frac{2N_\text{s}}{2-\alpha} \left(2-\alpha-e^{-\alpha t}  + (\alpha-1) e^{-2t}  \right).
\label{eqn:Average_m}
\end{eqnarray}
from which we calculate the average extent of reaction $ \langle p(t)
\rangle= {\langle m (t) \rangle} / {2 N_\text{s}}$
\begin{eqnarray}
\langle p (t) \rangle = 1 - \frac{1}{2-\alpha} [ e^{-\alpha t} - (\alpha -1) e^{-2 t}],
\label{eqn:poft} 
\end{eqnarray}
Inverting the transcendental Equations~\ref{eqn:Average_m}
and \ref{eqn:poft} for general $\alpha$ is not possible, 
however, under neutral cooperativity $\alpha=1$, we find
\begin{eqnarray}
\langle p(t) \rangle = 1 - e^{-t }.
\label{eqn:inverted_m}
\end{eqnarray}
A simple analysis of Equations~\ref{eqn:poft} and
\ref{eqn:inverted_m} reveals that $\langle p(t) \rangle$ is a
monotonically increasing function of $t$ for all $\alpha > 0$, which
is expected given that end-groups bind but do not unbind.  For $\alpha
=1$, Equations~\ref{eqn:MassActionSol} can be recast as
\begin{subequations}
\begin{align}
 \langle n_0  \rangle= & N_\text{s} (1- \langle p \rangle), \\
 \langle n_1  \rangle = & 2 N_\text{s}   \langle p \rangle (1- \langle p \rangle), \\
  \langle n_2  \rangle = &   N_\text{s}  \langle p \rangle^2. 
  \end{align}
  \label{eqn:MassActionSol2}
\end{subequations}
\noindent
Equations~\ref{eqn:MassActionSol2} have the same form as
Equation~\ref{eqn:avg}, obtained using mean-field arguments.  This
implies that the mean-field approach for a given extent of reaction
$p$ and $\alpha =1$ corresponds to an irreversible (quenched)
stochastic process halted at time $t^*$ such that $\langle p(t^*)
\rangle$ in Equation~\ref{eqn:inverted_m} satisfies $\langle p(t^*)
\rangle = 1 - e^{-t^*} = p$.  We plot the normalized average strand
numbers $\langle n_\ell(t) \rangle /N_\text{s}$ as a function of time
and as derived from Equations~\ref{eqn:MassActionSol}a-c in
Figure~\ref{fgr:Meq}a-c, for $N_\text{s}=40$ and different values of
$\alpha$. We find that $s_\text{1}$-strand formation is favored at
smaller $\alpha$, and $s_\text{2}$-strand formation is favored at
higher $\alpha$, as might be expected.  Since $\langle p(t) \rangle$
is a monotonic function of time we can plot $\langle n_\ell(t) \rangle
/N_\text{s}$ using Equations~\ref{eqn:MassActionSol}a-c as parametric
equations against $\langle p(t) \rangle$ given in
Equation~\ref{eqn:poft}.  Results are shown in Figure~\ref{fgr:Meq}d-f
for various values of $\alpha$.  These curves differ from those in
Figure~\ref{fgr:PartFunc}d obtained under equilibration and
calculated via Equations~\ref{eqn:AveragePartFunc}a-c.  Most
noticeably, $\langle n_1(t) \rangle$ loses its symmetry about $\langle
p(t) \rangle = 0.5$ and becomes skewed.

\begin{figure*}[t!]
\centering
\includegraphics[width=6in]{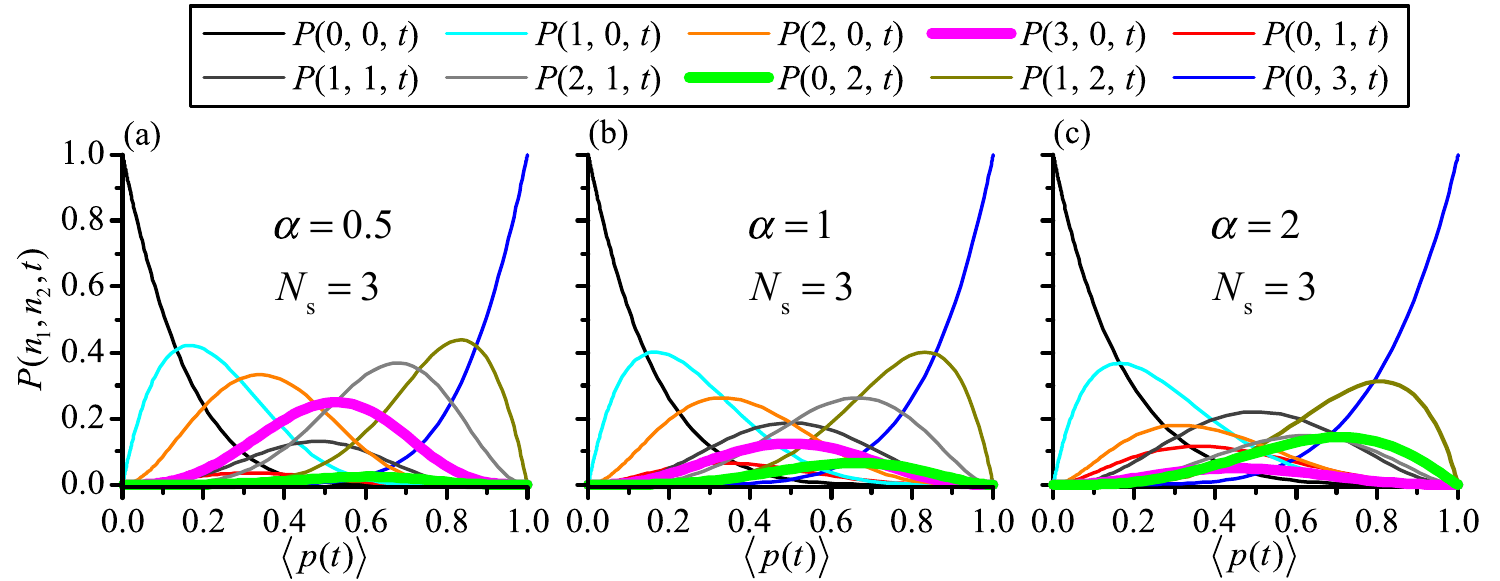}
        \caption{Probability distributions $P(n_1,n_2,t)$ for $N_{\rm
            s}=3$ under quenched binding as evaluated from
          Equation~\ref{eqn:MasterEqSol}.  We plot $P(n_1,n_2,t)$
          parametrically against $\langle p(t) \rangle$ as evaluated
          from Equation~\ref{eqn:poft} for \textbf{(a)} $\alpha=0.5$,
          \textbf{(b)} $\alpha=1$, \textbf{(c)} $\alpha=2$.  Of the
          ten possible configurations, two are highlighted: $\{ n_0,
          n_1, n_2\}=$ $\{ 0,3, 0\}$ (magenta) and $\{ 1, 0, 2\}$
          (green).}
\label{fgr:P_n1-n2_Ns3}
\end{figure*}
\begin{figure*}[t!]
\centering
\includegraphics[width=6.51 in]{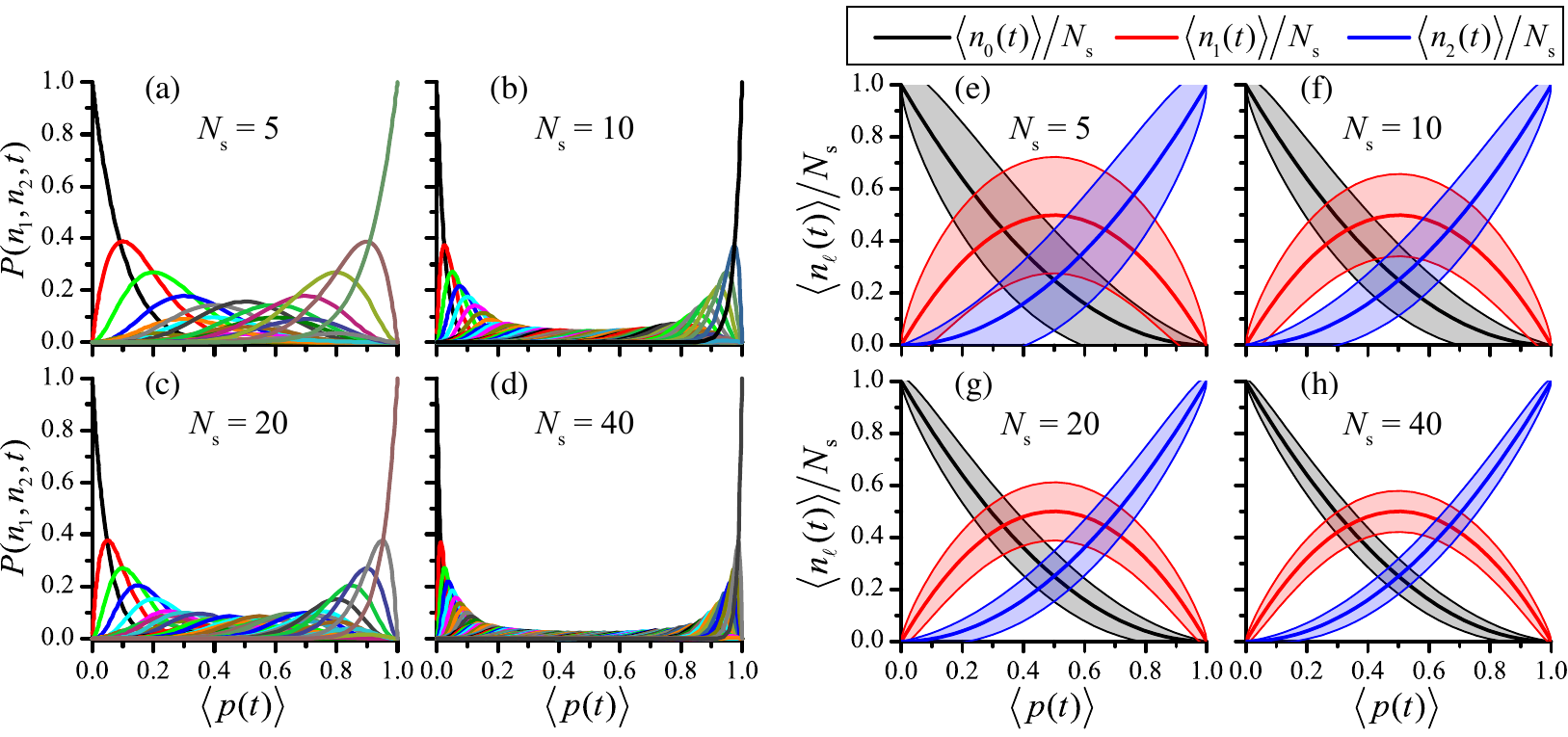}
      \caption{\textbf{(a-d)} Probability distributions $P(n_1,n_2,t)$
        for several values of $N_{\rm s}$ under quenched binding as
        evaluated from Equation~\ref{eqn:MasterEqSol}.  We set $\alpha
        =1 $ and plot $P(n_1,n_2,t)$ parametrically against $\langle
        p(t) \rangle$ as evaluated from Equation~\ref{eqn:poft}.
        \textbf{(e-h)} Average strand fractions $\langle
        n_\ell(t)\rangle / N_\text{s}$ for $\ell = 0,1,2$ plotted
        parametrically against $\langle p(t) \rangle$.  Standard
        deviations are calculated as the square root of the variance
        in Equation~\ref{eqn:MeqVariance}; shaded areas represent the
        associated error intervals. Chosen $N_{\rm s}$ values are
        \textbf{(a,e)} $N_\text{s}=5$, \textbf{(b,f)} $N_\text{s}=10$,
        \textbf{(c,g)} $N_\text{s}=20$, and \textbf{(d,h)}
        $N_\text{s}=40$.}
      \label{fgr:CombinedProbabilities2}
\end{figure*}

 
The master Equation~\ref{eqn:MasterEqQuench} also allows us to derive
the time-dependent likelihood of each of the many possible
configurations (enumerated in Equation~\ref{eqn:count}), a much more
powerful tool than average quantities.  For example,
Equation~\ref{eqn:MasterEqQuench} can be solved to find $P(n_1,
n_2,t)$ for all times $t$ once the initial condition is specified.  We
set this to be $P(n_1 =0, n_2=0 ,t=0) = 1$, and $P(n_1, n_2 ,t=0) = 0$
for all other values of $n_1, n_2 \neq 0$, so that the micro-region is
initially made only of free strands.  If one chooses to solve
Equation~\ref{eqn:MasterEqQuenchedHybrid2} to find $P(n_0, n_1, t)$
the equivalent initial conditions are $P(n_0 =N_{\rm s}, n_1=0 ,t=0) =
1$ and $P(n_0, n_1 ,t=0) = 0$ for $n_0 \neq N_{\rm s}$ and $n_1 \neq
0$.  We solve Equation~\ref{eqn:MasterEqQuenchedHybrid2} for $P(n_0,
n_1,t)$ rather than Equation~\ref{eqn:MasterEqQuench} for $P(n_1,
n_2,t)$ as the analytical computations are simpler.  From $P(n_0,
n_1,t)$ we can then derive $P(n_1, n_2,t)$ by changing variables
through the $n_0 + n_1 + n_2 = N_{\rm s}$ constraint.  To proceed, we
introduce the generating function $G(z_0,z_1,t)$ defined as
\begin{eqnarray}
\label{gen}
G(z_0,z_1,t) = \sum_{n_0, n_1} P(n_0, n_1, t) z_0^{n_0} z_1^{n_1},
\end{eqnarray}
\noindent
under the constraint $0 \leq n_0 + n_1 \leq N_\text{s}$.  Upon
multiplying Equation~\ref{eqn:MasterEqQuenchedHybrid2} by $z_0^{n_0}
z_1^{n_1}$ and summing over $n_0, n_1$, under the same constraint, we
find the following differential equation for $G(z_0,z_1,t)$

\begin{equation}
\label{eqn:char}
\frac{\partial G}{ \partial t} =  2 (z_1-z_0) \frac{\partial G}{\partial z_0} + \alpha (1- z_1) \frac{\partial G}{\partial  z_1}. 
\end{equation}
\noindent 
Equation~\ref{eqn:char} is coupled to the corresponding initial
condition $G(z_0,z_1,t=0)= z_0^{N_\text{s}}$.  Using the method of
characteristics, we find
\begin{equation}
\begin{split}
G(z_0,z_1,t) =& \Bigg[z_0 e^{- 2t} + \frac{2 z_1  \left(e^{-\alpha t} - e^{- 2t}\right)}{2-\alpha}  \\   
 &+1  + \frac{\alpha e^{-2t }-2e^{-\alpha t} }{2-\alpha}\Bigg]^{N_\text{s}}. 
 \end{split}
\end{equation}
After performing a Taylor series expansion in $z_0,z_1$ and upon
comparison with Equation\,\ref{gen} we find
\begin{equation}
\begin{split}
P(n_0, n_1, t) =& {N_\text{s} \choose {n_0, n_1}} e^{-2t n_0}
\left( \frac{2( e^{-\alpha t}-  e^{-2t}) }{2- \alpha} \right)^{n_1}  \\
  & \times \left(1 + \frac{\alpha e^{-2t} - 2e^{-\alpha t} }{2-\alpha} \right)^{N_{\rm s} - n_0 - n_1}.
\label{eqn:MasterEqSol0}
\end{split}
\end{equation}
From the constraint $n_0 = N_{\rm s} - n_1 - n_2$ we can finally write
\begin{equation}
\begin{split}
\label{eqn:MasterEqSol}
P(n_1, n_2, t) =& {N_\text{s} \choose {n_1, n_2}} e^{-2t (N_\text{s} - n_1 - n_2)} \\
 &\times \left( \frac{2  (e^{-\alpha t}-e^{-2 t} )}{2- \alpha} \right)^{n_1}  \\
 &\times \left( 1 + \frac { \alpha e^{-2t} - 2e^{-\alpha t} }{2-\alpha} \right)^{n_2}.  
\end{split}
\end{equation}
Note that $P(n_1, n_2, t \to \infty) = 0$ for $\{n_1, n_2 \} \neq \{0,
N_\text{s}\}$ and that $P(0, N_\text{s}, t \to \infty) = 1$ as
expected from a forward process.  Figure~\ref{fgr:P_n1-n2_Ns3} shows
the probability of individual configurations $P(n_1,n_2,t)$ of
micro-regions with $N_\text{s}=3$ plotted parametrically against
$\langle p(t) \rangle$ for different values of $\alpha$.  Two
different configurations are highlighted: $\{ n_0, n_1, n_2\}=$ $\{ 0,
3, 0\}$ and $\{ 1, 0, 2\}$.  For $\alpha$ = 2, when end-group binding
is cooperative, the probability of configurations with more
$s_\text{1}\text{-strands}$ decreases and those with more
$s_\text{2}\text{-strands}$ increases compared to the neutral ($\alpha
=1$) or uncooperative ($\alpha =0.5$) cases shown here.  In highly
cooperative scenarios, once a network strand has bound the transition
towards a fully-bound $s_2$-strand is fast.
In Figure~\ref{fgr:CombinedProbabilities2}a-d we plot the micro-region
configuration probabilities $P(n_1,n_2,t)$ as a function of the
average extent of reaction $\langle p(t) \rangle$ with increasing
$N_\text{s}$.  The highest value of $N_{\text s} =40$ we used results
in 861 distinct $\{ n_0, n_1, n_2\}$ micro-region configurations, as
per Equation~\ref{eqn:count}, all with non-zero probability at finite
time.  Larger values of $N_{\text s}$ are possible, but graphically
difficult to display.
 
By inserting the explicit expression for $P(n_1,n_2,t)$ from
Equation~\ref{eqn:MasterEqSol} into Equation~\ref{eqn:MassAction} we
evaluate the average values $\langle n_\ell(t)\rangle$ for $\ell =
1,2$ to reobtain the same expressions for $\langle n_\ell(t)\rangle$,
for $\ell = 0,1,2$ already displayed in
Equations~\ref{eqn:MassActionSol}.  From
Equation~\ref{eqn:MasterEqSol} we can also calculate the second
moments $\langle n^2_\ell(t)\rangle$ as
\begin{eqnarray}
\label{eqn:mom}
\langle n^2_\ell(t)\rangle = \sum_{n_1, n_2}  n_\ell^2 P(n_1,n_2,t)
\end{eqnarray}
for $\ell = 1,2$, and the correlation function
\begin{eqnarray}
\label{eqn:mommix}
\langle n_1(t) n_2(t) \rangle = \sum_{n_1, n_2}  n_1 n_2  P(n_1,n_2,t)
\end{eqnarray}
from which we can derive $\langle m^2 (t) \rangle$
\begin{equation}
\begin{split}
\langle m^2(t)\rangle =&  \sum_{n_1, n_2}  (n_1 + 2 n_2) ^2 P(n_1,n_2,t)  \\
=& \langle n^2_1(t)\rangle + 4 \langle n^2_2(t)\rangle + 4 \langle n_1(t) n_2(t) \rangle.
\label{eqn:m2}
\end{split}
\end{equation}
Using Equations~\ref{eqn:MassActionSol}, \ref{eqn:Average_m},  \ref{eqn:mom},
\ref{eqn:mommix} we find the variances 
\begin{equation}
\begin{split}
\mathrm{Var}\big(n_\ell(t)\big) =& \langle n^2_\ell(t)
\rangle -  \langle n_\ell(t)\rangle^2,  \\
\mathrm{Var}\big(m(t)\big) =& \langle m^2(t) \rangle -  \langle m(t)\rangle^2, 
\label{eqn:MeqVariance}
\end{split}
\end{equation}
for $\ell = 1,2$. Finally, $\langle n^2_0 \rangle$ is obtained as
\begin{equation}
\begin{split}
\langle n^2_0(t)\rangle =& \sum_{n_1, n_2}  (N_{\rm s} - n_1 - n_2)^2 P(n_1,n_2,t)  \\
 =& N^2_{\rm s} + \langle n^2_1(t)\rangle +  \langle n^2_2(t)\rangle + N_{\rm s}  \langle n_1(t)\rangle   \\
&+ N_{\rm s} \langle n_2(t)\rangle + \langle n_1(t) n_2(t) \rangle.
\label{eqn:mom0}
\end{split}
\end{equation}

Equations~\ref{eqn:mom0} and \ref{eqn:MassActionSol} yield
$\mathrm{Var}\big(n_0(t)\big) = \langle n^2_0(t) \rangle - \langle
n_0(t)\rangle^2$. Explicit expressions for $\langle n^2 _{\ell} (t)
\rangle$, $\langle m^2 (t) \rangle$,
$\mathrm{Var}\big(n_{\ell}(t)\big) $ and $\mathrm{Var}\big(m(t)\big) $
are presented in Section \ref{appendix:sm} of the Appendix. In
Figure~\ref{fgr:CombinedProbabilities2}e-h we show the parametric
plots of the average strand fractions $\langle n_{\ell} (t) \rangle /
N_{\rm s}$ for $\ell = 0,1,2$ against the average extent of reaction
$\langle p(t) \rangle$ for several values of $N_{\rm s}$.  The
associated standard deviations calculated as the square root of the
variance in Equation~\ref{eqn:MeqVariance} are also displayed.  As
can be expected, fluctuations decrease as $N_{\rm s}$ increases.
\begin{figure}[t!]
\centering
\includegraphics[width=3.375 in]{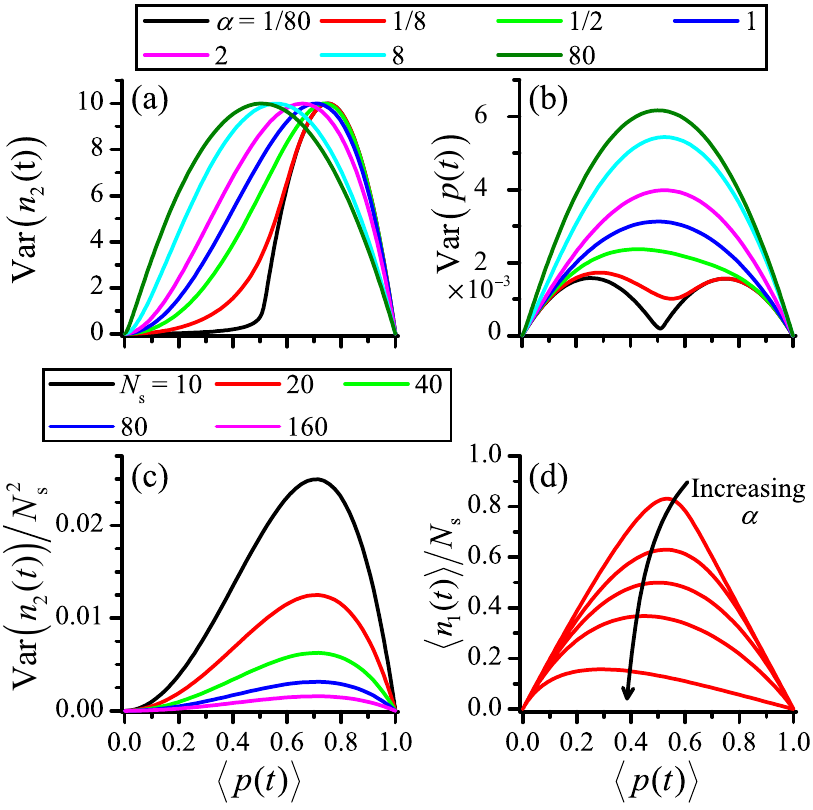}
\caption{\textbf{} The variance of the number of \textbf{(a)}
  $s_\text{2}\text{-strands}$ and of \textbf{(b)} bound end-groups $m$
  as evaluated by Equations~\ref{eqn:MeqVariance} for $\alpha$= 1/8,
  1/2, 1, 2, or 8 and $N_\text{s}=40$.  $\mathrm{Var}(n_2(t))$ and
  $\mathrm{Var}(p(t))$ are plotted parametrically against $\langle
  p(t) \rangle$ as evaluated from Equation~\ref{eqn:poft}.  For
  $\alpha>80$ and $\alpha<1/80$ the curves do not change significantly
  from those displayed.  \textbf{(c)} The variance of the fraction of
  $s_\text{2}\text{-strands}$ $\mathrm{Var}(n_2)/N^2_\text{s}$ for
  $N_\text{s}=10,20,40,80,160$ and $\alpha = 1$.  \textbf{(d)} The
  relative $\langle n_1(t)\rangle/N_\text{s}$ as calculated from
  Equation~\ref{eqn:MassActionSol}b for $\alpha$= 1/8, 1/2, 1, 2, or 8
  and $N_\text{s}=40$.}
\label{fgr:MeqVarCombined}
\end{figure}
Figure~\ref{fgr:MeqVarCombined}a shows the parametric plot of
$\mathrm{Var}\big(n_2(t)\big)$ against $\langle p(t) \rangle$ for
different values of $\alpha$ and $N_\text{s} = 40$.  For strong uncooperative
binding ($\alpha \to 0$) and small extents of reaction $\langle p(t)
\rangle$, only free strands bind to the network and the variance is
very small. However, once all strands have bound at least at one end,
and $\langle p(t) \rangle \sim 0.5$, the dangling strands transition
to the fully bound state and the variance starts increasing. In
Equation~\ref{eqn:maxvar} of Section~\ref{appendix:sm} of the Appendix we
give an exact analytical expression for
$\mathrm{Var}\big(n_2(t)\big)$; a simple calculation shows that the
maximum variance is $N_{\rm s}/4$ for all values of $\alpha$, and is
attained at smaller average extents of reaction $\langle p(t) \rangle$
as $\alpha$ increases.  In Figure~\ref{fgr:MeqVarCombined}b we plot
the parametric dependence of $\mathrm{Var}\big(p(t)\big)$ against the
average extent of reaction $\langle p(t) \rangle$ with variable
$\alpha$ and $N_\text{s}=40.$ For very small values of $\alpha \to 0$,
$\mathrm{Var}\big(p(t)\big)$ is bimodal and approximately zero at
$\langle p(t) \rangle \to 0.5$.  This is because, as discussed above,
for $\alpha \to 0$ end-group binding occurs only on free strands for
$\langle p(t) \rangle < 0.5$ and the most likely configurations are
those with $s_0$ and $s_1$ strands.  As $\langle p(t) \rangle \to
0.5$, only $s_1$ dangling strands remain so that $\langle n_1(t)
\rangle \to N_{\rm s}$, $\langle n^2_1(t) \rangle \to N^2_{\rm s}$,
$\mathrm{Var}\big(n_2(t)\big) \to 0$ and as a result,
$\mathrm{Var}\big(p(t)\big) \to 0$.  Fully bound strands start
emerging for $\langle p(t) \rangle > 0.5$, increasing
$\mathrm{Var}\big(p(t)\big)$. As $\alpha$ increases the variance
increases for all $\langle p(t) \rangle$ and the minimum at $\langle
p(t) \rangle \sim 0.5$ turns into a maximum.  A more detailed
discussion is presented in Section~\ref{appendix:sub} of the Appendix.

In Figure~\ref{fgr:MeqVarCombined}c we plot $\mathrm{Var}\big(n_2(t)
\big)/N^2_{\rm s}$ against $\langle p(t) \rangle$ for different values
of $N_\text{s}$; the curves decrease in magnitude as $N_{\rm s}$
increases.  Finally, in Figure~\ref{fgr:MeqVarCombined}d we plot
$\langle n _1(t) \rangle / N_{\rm s}$ parametrically against $\langle
p(t) \rangle$ for several values of $\alpha$.  The curves decrease
with $\alpha$ once $\langle p(t) \rangle$ is fixed. This also follows
from Equation~\ref{eqn:MassActionSol}b which implies that $\langle n
_1(t) \rangle $ is a decreasing function of $\alpha$ for any time $t$.
Since $\langle p(t) \rangle$ is univocally associated to $t$ via
Equation~\ref{eqn:poft} it also follows that $\langle n _1(t) \rangle
$ is a decreasing function of $\alpha$ for any $\langle p(t) \rangle$.
Equations~\ref{eqn:MassActionSol}b and \ref{eqn:poft} reveal that the
maximum $\langle n _1(t_{\rm max}) \rangle / N_{\rm s} =
(\alpha/2)^{(1-2/\alpha)}$ is attained at $\langle p (t_{\rm max})
\rangle = 1 - ((\alpha + 1)/\alpha) e^{-2 t_{\rm max} }$ where $t_{\rm
  max} = \ln (2 /\alpha)/(2 - \alpha)$. One can easily verify that
$\langle p (t_{\rm max}) \rangle$ is a decreasing function of $\alpha$
as well.  These results can be expected as larger $\alpha$ favors the
formation of fully bound strands. Thus, for a given average extent of
reaction $\langle p(t) \rangle$ the fraction of dangling ends
decreases with $\alpha$, and the maximum is found at an average extent
of reaction $\langle p(t) \rangle$ that also decreases with $\alpha$.

\subsubsection{Dynamic end-group rearrangement/redistribution}
\label{dynamic}
We now consider an equilibration process that allows the bound
end-groups in a micro-region to dynamically rearrange, attaching and
detaching until thermodynamic equilibrium is reached\cite{Bowman2012}
while maintaining a fixed total number of $m$ bound end-groups.  We
assume that $m<2 N_\text{s}, (p < 1)$ so that the reaction is not
complete and multiple $\{n_0, n_1, n_2 \}$ configurations are
possible. Experimental realizations include the formation reversible
hydrazone bonds \cite{Roberts2007}, imine bonds \cite{Yang2012}, or
guest-host interactions \cite{Rodell2015}.  Here, an intact,
$s_\text{2}\text{-}$strand may detach at one of its ends to form a
dangling $s_\text{1}\text{-}$strand, while a free
$s_\text{0}\text{-}$strand binds to the network to form another
$s_\text{1}\text{-}$strand.  The reverse process where two
$s_\text{1}\text{-}$strands become an $s_\text{2}\text{-}$ and
$s_\text{0}\text{-}$strand is also possible.  In all scenarios $m =
n_1 + 2n_2$ is fixed, but there are many distinct ways for the bound
end-groups to distribute in $s_\text{1}\text{-}$ or
$s_\text{2}\text{-}$ strands.  The final
equilibrium configuration is independent of
initial conditions so our results will depend only on the selected
value of $m$. We write the reversible master
equation for $P(n_1,n_2,t)$ as
\begin{equation}
\begin{split}
\frac{\text{d}P(n_1, n_2, t)}{\text{d}t} =& 
2\kappa \alpha^2 { n_1 +2 \choose 2 } P(n_1+2, n_2 -1,t)  \\
& + 4 \kappa  (N_\text{s}-n_1-n_2+1) (n_2 +1)  \\
& \times P(n_1-2, n_2+1, t)   \\
& - 2  \kappa  \alpha^2{ n_1  \choose 2 } P(n_1, n_2, t)  \\
& - 4\kappa n_2(N_\text{s}-n_1-n_2)   P(n_1, n_2, t).
\label{eqn:Hopping}
\end{split}
\end{equation}
Here, $\kappa$ is dimensionless and represents the rearrangement rate 
measured in terms of the binding rate.
The first term on the right hand side of Equation~\ref{eqn:Hopping} accounts for the
formation of a fully bound $s_\text{2}\text{-strand}$ and a free
$s_\text{0}\text{-strand}$ from two dangling
$s_\text{1}\text{-strands}$ ($2s_\text{1} \rightarrow
s_\text{0}+s_\text{2}$).  Here, the bound end-group of one of the two
$s_\text{1}\text{-strands}$ exchanges with the unbound end-group of
the other leading to the $\{n_0-1, n_1 + 2 , n_2 -1 \} \to \{n_0, n_1,
n_2\}$ transition.  The combinatorial factor enumerates the number of
$s_\text{1}\text{-strands}$ present in the micro-region and the 2
prefactor represents both $s_\text{1}\text{-strands}$ being able to
exchange with the other.  Finally, the reactivity parameter $\alpha$
is squared, since two dangling ends must bind to form a fully bound
strand.  The second term represents a fully bound
$s_\text{2}\text{-strand}$ detaching on one end while promoting the
binding of a free $s_\text{0}\text{-strand}$, giving rise to two
dangling $s_\text{1}\text{-strands}$. This process is represented by
the $\{n_0+1, n_1-2, n_2 +1 \} \to \{n_0, n_1, n_2 \}$ transition.
The factors $(N_\text{s}-n_1-n_2+1)$ and $(n_2+1)$ represent the number of
$s_\text{0}\text{-}$ and $s_\text{2}\text{-}$network strands
available, respectively.  The prefactor 4 accounts for the number of
possible bond movements: either of the two bound end-groups on the
$s_\text{2}$-strand can relocate to either of the two unbound
end-groups of the $s_\text{0}$-strand, yielding a total of four
combinations. The last two terms represent the same two processes
described above, but driving the system away from $\{n_0, n_1, n_2
\}$.  Note that there are no terms in Equation~\ref{eqn:Hopping}
representing bonds leaving an $s_\text{2}$-strand to populate an
$s_\text{1}$-strand; this transition would not change the overall the
micro-region configuration $\{n_0, n_1, n_2 \}$.  The probability
$P_\text{b}(m,t)$ of having $m$ bound-ends at time $t$ can be written
as

\begin{eqnarray}
P_\text{b}(m,t) = \sum_{n_2=0}^{[ m/2 ]} P(m - 2 n_2, n_2, t),
\end{eqnarray}

\noindent
where all possible $n_1, n_2$ combinations that yield $m= n_1 + 2 n_2$
are included.  Using Equation~\ref{eqn:Hopping} it can be easily
verified that $P_\text{b} (m,t)= P_{\rm b}(m,t=0)$. As expected, the
master Equation~\ref{eqn:Hopping} only rearranges the distribution of
$s_\text{1}$ and $s_\text{2}$ strands but $m$ remains unchanged.  We
thus assume the system is initiated with a given $m$ so that $n_1+2n_2
= m$ at all times.  In addition to this constraint, the number of
strands is fixed so that $n_0+n_1+n_2=N_\text{s}$.  We can thus cast
Equation~\ref{eqn:Hopping} in terms of only one of the $n_0,n_1$ or
$n_2$ populations.  We choose $n_2$ and determine the steady state
$P(n_1, n_2, t \to \infty) \equiv P^*(n_2)$ by imposing detailed
balance between the first and the last term on the right hand side of
Equation~\ref{eqn:Hopping}, or equivalently, the second and the third,
since it can be easily verified that the conditions are the same. We
find
\begin{equation}
\begin{split}
\label{eqn:HoppingSol0}
\frac{P^*(n_2-1)}{P^*(n_2)} = & \frac{4 n_2(N_\text{s} - m +n_2)}{\alpha^2 (m - 2n_2 + 2)(m - 2 n_2+1)},
\end{split}
 \end{equation}
which can be solved to yield
\begin{equation}
\label{eqn:HoppingSol}
P^*(n_2) = \frac{1}{Z_{m,N_\text{s}}} \frac{(2/\alpha)^{m - 2 n_2} 
N_\text{s}!}{(m-2n_2)! n_2! (N_\text{s} - m + n_2)!},
\end{equation}
where $Z_{m,N_\text{s}} $ is the normalization constant
\begin{equation}
Z_{m,N_\text{s}} = \sum_{n_2=0}^{[m/2]} \frac{(2/\alpha)^{m - 2 n_2}  
N_\text{s}!}{(m-2n_2)! n_2! (N_\text{s} - m + n_2)!}.
\end{equation}
\noindent
This result is the same as Equation~\ref{eqn:bias}: the combinatoric
approach for a fixed $p$ is equivalent to allowing for relaxation on
the network with a fixed number of bound ends and $m = 2 p N_{\rm s}$.

\subsubsection{End-group rearrangement/redistribution and bond formation}
We now consider the two processes of bond formation and redistribution
occurring simultaneously and combine the two master
Equations~\ref{eqn:MasterEqQuench} and \ref{eqn:HoppingSol} so that
\begin{equation}
\begin{split}
\frac{\text{d}P(n_1, n_2, t)}{\text{d}t} 
 = & 2  (N_\text{s} - n_1 - n_2 +1) P(n_1 - 1, n_2,t) \\
& + \alpha (n_1 +1) P(n_1 +1, n_2 - 1,t) \\
& + \kappa \alpha^2 (n_1 +2) (n_1 +1) \\
& \times P(n_1+2, n_2 -1,t) \\
& + 4 \kappa  (N_\text{s}-n_1-n_2+1) (n_2 +1) \\
&\times P(n_1-2, n_2+1, t) \\
& - \left[\kappa  \alpha^2  n_1 (n_1 -1) +  4\kappa n_2(N_\text{s}-n_1-n_2) \right. \\ 
& + \left. 2 (N_\text{s}  - n_1 - n_2) + \alpha  n_1\right] P(n_1, n_2,t).
\end{split}
\label{eqn:Hopping2}
\end{equation}
Fast annealing, fast binding, and quenched/irreversible binding are modeled by 
setting $\kappa \gg 1$, $\kappa \ll 1$, and $\kappa = 0$, respectively.
\begin{figure}[t!]
\centering
\includegraphics[width=3.375 in]{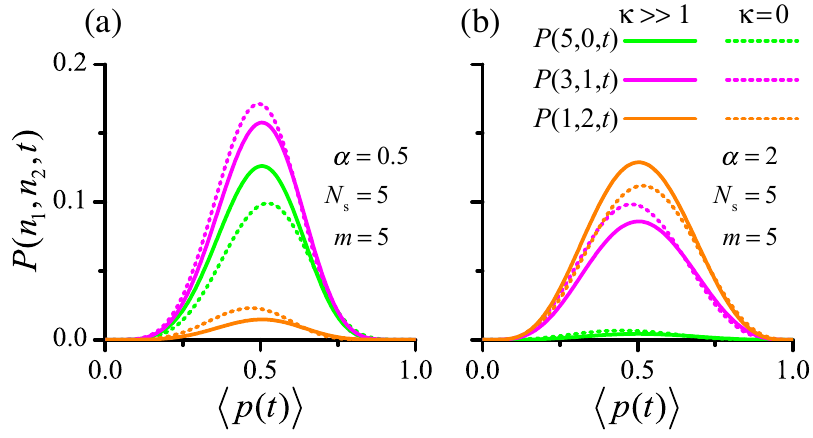}
\caption{Configuration probabilities $P(n_1,n_2,t)$ 
calculated from Equation~\ref{eqn:Hopping2} and plotted
  parametrically against $\langle p(t) \rangle$ under fast annealing
  ($\kappa=1000$, solid lines) and quenched binding
  ($\kappa=0$, dashed lines) for $N_{\rm s} = 5$ and
  \textbf{(a)} $\alpha = 0.5$, $m=5$; \textbf{(b)} $\alpha = 2$,
  $m=5$.}
\label{fgr:HoppingMeqVariance}
\end{figure}
Although a full analytical time-dependent solution can not be
  found, the effects of annealing can be observed in
Figure~\ref{fgr:HoppingMeqVariance}.  Here we parametrically plot
$P(n_1, n_2,t)$ against $\langle p(t) \rangle$ using
Equation~\ref{eqn:Hopping2} for $N_{\rm s} = m = 5$, $\alpha=0.5$ and
$\alpha =2$, under fast annealing ($\kappa = 1000$) or
quenched binding ($\kappa = 0$).  Since the rearrangement
process allows for more configurations to be explored we expect
cooperative effects to be more pronounced under fast annealing, than
under quenched binding.  In Figure~\ref{fgr:HoppingMeqVariance}a we
set $\alpha =0.5$; since binding is uncooperative, annealing favors
configurations with lower values of $n_2$. Indeed, the $k = 1000$ curves show an increase in $P(5,0,t)$ compared to the
corresponding $\kappa = 0$ curves, whereas $P(3,1,t),
P(1,2, t)$ decrease.  Similar trends are observed in
Figure~\ref{fgr:HoppingMeqVariance}b, where we set $\alpha =
2$. Cooperative binding increases the likelihood of configurations
with higher $n_2$, so in this case $P(1,2,t)$ increases while
$P(3,1,t), P(5,0,t)$ decrease.  Note that $P(5,0,t)$, $P(3,1,t)$, and
$P(1,2,t)$ all obey the constraint $n_1 + 2 n_2 = N_{\rm s} = 5$.  For
$\alpha = 2$, and under quenched binding at $\kappa = 0$, Equation~\ref{eqn:MasterEqSol} yields $P(5,0,t) = 32 e^{-10 t} t^5$ which is
maximized at $t=1/2$ corresponding to $\langle p (t=1/2) \rangle = 1 -
3/2 e \neq 1/2$, as per Equation~\ref{eqn:poft}.  Similarly, $P(3,1,t)
= 8 e^{-8 t} t^3 (1-(2t +1) e^{-2 t})$ and $P(1,2,t) = 2 e^{-6 t} t
(1-(2t +1) e^{-2 t})$ are also maximized at times that correspond to
$\langle p(t) \rangle \neq 1/2$.  None of the three distribution
curves are thus symmetric about $\langle p(t) \rangle = 1/2$.  When
$\kappa = 1000$ however the master
Equation~\ref{eqn:Hopping2} yields numerical results that are closely
aligned with those derived from Equation~\ref{eqn:Hopping} upon
setting $n_1 + 2n_2 = N_{\rm s} = 5$. This is because annealing is
much faster than binding and the time between binding events is much
longer than the time for equilibration of a fixed number of bound
strands.  As a result, once a strand binds, the network can almost
fully equilibrate before the next binding event occurs.  The curves in
Figure~\ref{fgr:HoppingMeqVariance}b for $\kappa=1000$
thus mirror Equation~\ref{eqn:HoppingSol}, with the proportions
$P(5,0)$:$P(3,1)$:$P(1,2)$ following Equation~\ref{eqn:HoppingSol0}
and become symmetric about $\langle p(t) \rangle = 0.5$ as predicted
by Equation~\ref{eqn:HoppingSol} when $m = N_{\rm s}$.  The same
trends arise when comparing the quenched binding and the fast
annealing curves for the uncooperative ($\alpha=0.5$) case. 

\begin{figure}[t!]
\centering
\includegraphics[width=3.375 in]{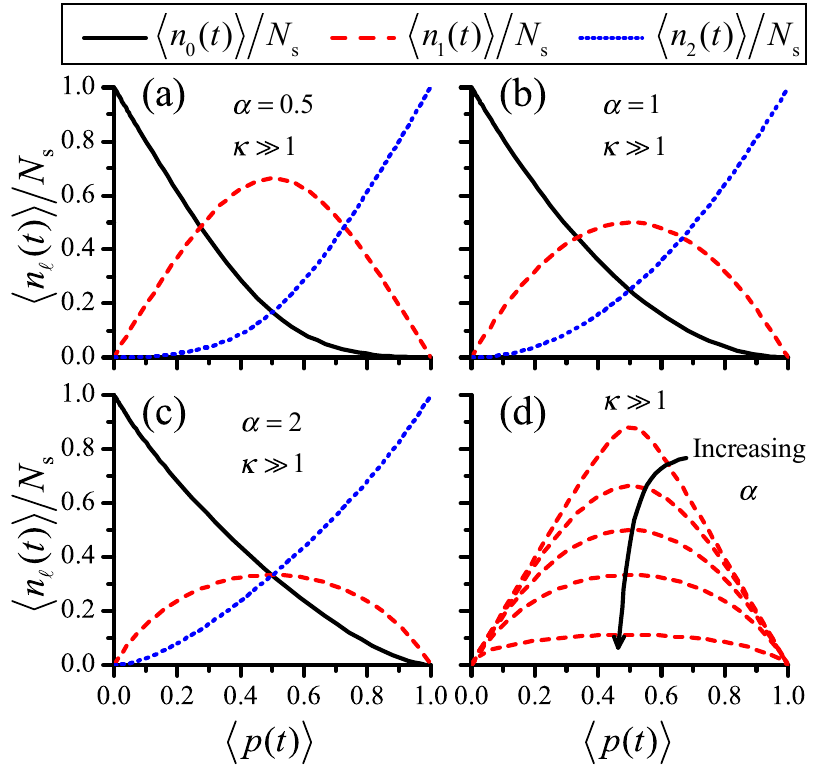}
\caption{\textbf{}{Average strand fractions $\langle n_\ell
    (t)\rangle/N_\text{s}$ for $\ell = 0,1,2$ and $N_{\rm s} =10$
    evaluated using the probability distribution in Equation~\ref{eqn:Hopping2} and plotted
    parametrically against $\langle p (t) \rangle$ under fast
    annealing ($\kappa=1000$).  The reactivity
    parameter is set as \textbf{(a)} $\alpha = 1$, \textbf{(b)}
    $\alpha = 0.5$, and \textbf{(c)} $\alpha = 2$.  All curves closely
    resemble those obtained from the equilibrium distribution
    in Equation~\ref{eqn:HoppingSol}.  \textbf{(d)} Under fast
    annealing $\langle n_1 (t)\rangle/N_\text{s}$ is symmetric about
    $\langle p(t) \rangle=0.5$ for all $\alpha$, here set at $\alpha=$
    1/8, 1/2, 1, 2, 8, from top to bottom. }}
      \label{fgr:HoppingMeq}
\end{figure}

In Figure~\ref{fgr:HoppingMeq} we plot $\langle
n_{\ell}(t)\rangle/N_\text{s}$ parametrically against $\langle p(t)
\rangle$ for $\alpha =$ 0.5, 1, 2 using the probability distribution
in Equation~\ref{eqn:Hopping2} for $\ell = 0,1,2$ and $\kappa = 1000$. In all cases, the solutions closely match those obtained from
the equilibrated distribution in Equation~\ref{eqn:HoppingSol} for
all values of $\alpha$ as can be seen upon comparison with
Figure~\ref{fgr:PartFunc}d.  The most notable feature is the
symmetry of $\langle n_1 (t)\rangle$ about $\langle p(t) \rangle =
0.5$ for all $\alpha$, a feature of the combinatoric approach, as
discussed in Section \ref{Sec:Combinatoric}, Intermediate values of
$\kappa \approx 1$ yield curves that interpolate between the two
extremes $\kappa \gg 1$ and $\kappa =0$ shown here.  Our results
imply that networks formed via quenched end-group binding, as per
Equation~\ref{eqn:MasterEqSol}, should not be described by models that
assume network strands equilibration via redistribution, as per
Equations~\ref{eqn:binomial} and \ref{eqn:pbias}.



\section{Applications and Conclusions}

In this work we studied the stochastic properties of bifunctional
network strands that undergo an end-linking gelation process.  We
developed and analyzed a master equation to describe quenched and
annealed binding events in micro-regions within a larger polymer
network, and include a reactivity parameter to model cooperative
effects.  While typical models quantify ``average'' mean-field
properties, we are able to evaluate the full probability distribution
for any given configuration as a function of time and extent of
reaction. By modeling the probability of a configuration within a
micro-region, we can propose a crude framework to describe the effects
of heterogeneity across the entire sample.

For example, our approach can be used in several polymer
network applications under the assumption that a macroscopic region is
comprised of a collection of statistically identical,
independent, smaller micro-regions.  For
example, nano/micrometer scale differences in the polymer network
properties can affect the fate of cells that are cultured on them
\cite{Yang2016} as well as the mechanical properties of
high-performance materials \cite{Li2016}. If these properties depend
on the local number of free, dangling, and intact strands, we can use
the relevant probability distribution to evaluate the likelihood of a
given configuration $\{n_0,n_1,n_2\}$ in any number of micro-regions
sampled by \textit{e.g.}, a cell. The statistical distribution for
each micro-region can then be used to construct the probability
distribution of the entire macro-system and thus to estimate the
chemical or mechanical properties of the polymer network, including
their local variability.

Similar considerations can be applied to the study of elastic
properties, in particular within phantom network theory which posits
that the shear modulus of an ideal network depends on the number
density of elastically effective network strands. Our results are
readily applicable if we assume that all $s_\text{2}\text{-strands}$
in our models are elastically effective and the number of branchpoints
is fixed. Starting from the probability $P(n_1, n_2, t)$
that a micro-region is in the $\{n_0, n_1, n_2 \}$ configuration, we can also compute the
likelihood that a given threshold is met, say, $n_{2} \geq
n_{2}^{*}$.  This quantity can then be interpreted as the probability for a ``bond"  to
stretch across a micro-region. One can then calculate the likelihood that
a given number of contiguous micro-regions with
$n_{2} \geq n_{2}^{*}$ span  the sample through percolation,
leading to a dramatic stiffening of the network. 

Finally, our work can also be applied to the study of network
degradation, which has attracted much attention over the past two
decades as degradable sites have been increasingly incorporated into
experimental realizations of end-linking polymerization. These
degradable strands typically cleave by enzymatic, hydrolytic,
photolytic, or other chemical mechanisms and allow for a reverse
gelation process.  Here, strands initially exist in the fully bound,
intact state where both ends are unreacted.  Reverse gelation occurs
via reaction or degradation of either end, so that intact strands
first become dangling strands, and dangling strands then become free
strands.  Halting the extent of reaction is common in degradable
networks as a way to tune the gel mechanics and this results in a
large variability of the micro-region composition.  Photodegradable
networks \cite{Griffin2012, Norris2017, Xue2014, Kapyla2016,
  Norris2019, Wong2010}, where end-groups are degraded by exposure to
light, are of particular interest as they are uniquely suited to
spatially pattern network stiffness, with a high degree of control
\cite{Norris2016}.  Some mathematical models of reverse gelation have
been formulated by adapting models of gelation \cite{Metters2000a};
more specific mean-field photodegradable network models have also been
proposed \cite{Tibbitt2013b, Norris2017}.  The present work can be
adapted to model degradable networks by associating intact network
strands to $s_\text{0}\text{-strands}$ (0 degraded end-groups),
dangling strands to $s_\text{1}\text{-strands}$ (1 degraded
end-group), and free strands to $s_\text{2}\text{-strands}$ (2
degraded end-groups). Cooperative effects arise in this context
  as the un-degraded end-groups of an intact strand might more readily
  react due to tension across the strand induced by network swelling.
  Once one of the end-groups has cleaved, and the strand dangles, the
  stress is removed so that the remaining un-degraded end-group is
  less susceptible to further degradation.  Using our stochastic
  framework, one can calculate the probability of any given
  micro-region configuration, distinguishing between quenched and
  annealed network de-gelation reactions. Melting and collapse of
  rigidity can be then be described using percolation concepts.
  
  \begin{acknowledgments}
Funding for this work was provided by the National Institutes of
Health through the NIH Director's New Innovator Award Program,
1-DP2-OD008533. S.C.P.N gratefully acknowledges support from a Ruth
L. Kirschstein Predoctoral Fellowship
(NIH-F31DE026356). T. C. acknowledges support from the NSF through
grant DMS-1814364.  M.R.D. acknowledges support from the NSF through
grant DMS-1814090 and Army Research Office W911NF-16-1-0165 . Both T.C. and M.R.D. also acknowledge support from
the Army Research Office (W911NF-18-1-0345).
\end{acknowledgments}


\appendix
\section{Second moments}
\label{appendix:sm}
\noindent
We here derive $\langle n^2 _{\ell} (t) \rangle$, $\langle m^2 (t)
\rangle$, $\mathrm{Var}\big(n_{\ell}(t)\big) $ and
$\mathrm{Var}\big(m(t)\big) $ for $\ell = 0, 1,2$ using the explicit
form for $P(n_1, n_2,t)$ as given in Equation~\ref{eqn:MasterEqSol}.
We begin with
\begin{eqnarray}
\label{eqn:n2l}
\langle n^2_{\ell}(t) \rangle = \sum_{n_2=0}^{N_{\rm s}} 
\sum_{n_1 = 0}^{N_{\rm s}- n_2} n^2_{\ell} P(n_1, n_2,t)
\end{eqnarray}
\noindent
and the associated variances for $\ell=1,2$. Using the binomial theorem we find
\begin{eqnarray}
\label{eqn:n21}
\langle n^2_1(t) \rangle &=& \frac{2 N_{\rm s}}{(2 - \alpha)^2} (e^{-\alpha t } - e^{-2 t}) \nonumber \\
&& \times [2-\alpha + 2 (N_{\rm s} -1) (e^{-\alpha t } - e^{-2 t})]
\end{eqnarray}
which, coupled with Equation~\ref{eqn:MassActionSol}b for $\langle n_1 (t) \rangle$ leads to 
\begin{eqnarray}
\mathrm{Var}\big(n_1(t)\big) &=& \frac{2 N_{\rm s}}{(2 - \alpha)^2} (e^{-\alpha t } - e^{-2 t}) \nonumber \\
&& \times (2-\alpha - 2 e^{-\alpha t } +2  e^{-2 t}).
\end{eqnarray}
Similarly, Equation~\ref{eqn:n2l} for $\ell=2$ yields
\begin{eqnarray}
\label{eqn:n22}
\langle n^2_2(t) \rangle &=& \frac{N_{\rm s}}{(2 - \alpha)^2} 
(2 - \alpha - 2 e^{-\alpha t } + \alpha e^{-2 t}) \nonumber \\
&& \times[N_{\rm s} (2 -\alpha -2 e^{-\alpha t } + \alpha e^{-2 t}) +
2 e^{-\alpha t } - \alpha e^{-2 t}], \nonumber \\
&&
\end{eqnarray}
which coupled with Equation~\ref{eqn:MassActionSol}c for $\langle n_2 (t) \rangle$ leads to
\begin{eqnarray}
\label{eqn:maxvar} 
\mathrm{Var}\big(n_2(t)\big) &=& \frac{ N_{\rm s}}{(2 - \alpha)^2} 
(2-\alpha - 2 e^{-\alpha t } +\alpha  e^{-2 t}) \nonumber \\
&& \times (2 e^{-\alpha t } - \alpha e^{-2 t}).
\end{eqnarray}
Equation~\ref{eqn:maxvar} is maximized for time $t_{\rm M}$ implicitly given by
\begin{eqnarray}
2 e^{-\alpha t_{\rm M}}- \alpha e^{-2 t_{\rm M}} = \frac{2-\alpha}{2}
\end{eqnarray}
which corresponds to $\mathrm{Var}\big(n_2(t_{\rm M})\big) = N_{\rm s}/4$,
independent of the value of $\alpha$. 
To evaluate $\langle m^2 (t) \rangle$ we must 
first calculate the correlation
function
\begin{eqnarray}
\langle n_1(t) n_2(t)\rangle = \sum_{n_2=0}^{N_{\rm s}} \sum_{n_1 = 0}^{N_{\rm s}- n_2} n_1 n_2 P(n_1, n_2,t)
\end{eqnarray}
which yields
\begin{eqnarray}
\langle n_1(t) n_2(t)\rangle &=& 
\frac{2 N_{\rm s} (N_{\rm s} - 1)}{(2 - \alpha)^2}
 (e^{-\alpha t } - e^{-2 t}) \nonumber \\
 && \times (2-\alpha - 2 e^{-\alpha t } +\alpha  e^{-2 t}) 
\end{eqnarray}
We can now evaluate $\langle m^2(t)\rangle$ using
Equation~\ref{eqn:m2}, \ref{eqn:n21} and \ref{eqn:n22}
\begin{eqnarray}
\langle m^2(t)\rangle  &= &
\frac{4 N_{\rm s}^2}{(2 - \alpha)^2} 
\left[2 - \alpha+ (\alpha -1) e^{-2t} - e^{-\alpha t}\right]^2 \nonumber \\
& & +  \frac{2 N_{\rm s}}{(2 -\alpha)^2}
\Big[(2 - \alpha)(e^{-\alpha t} + (3 - 2 \alpha) e^{-2 t})  \nonumber \\
&&-  2 \left((\alpha -1)e^{-2t} -e ^{-\alpha  t}\right)^2 \Big], 
\end{eqnarray}
from which the variance is obtained as
\begin{eqnarray}
\mathrm{Var}\big(m(t)\big) &=& 
 \frac{2 N_{\rm s}}{(2 -\alpha)^2}
\Big[(2 - \alpha)\left(e^{-\alpha t} + (3 - 2 \alpha) e^{-2 t}\right) \nonumber \\ 
&& -2 \left[(\alpha -1)e^{-2t} -e ^{-\alpha  t}\right]^2 \Big],
\label{eqn:varm}
\end{eqnarray}
where we used Equations~\ref{eqn:MassActionSol}b-c to evaluate $\langle m(t) \rangle$.
Finally, using the constraint $n_0 = N_{\rm s} - n_1 -n_2$ we find
\begin{equation}
\langle n^2_0(t) \rangle =
N_{\rm s} \left[N_{\rm s} e^{-4 t } + e^{-2 t} (1- e^{-2 t})\right], 
\end{equation}
which together with Equation~\ref{eqn:MassActionSol}a
finally yields
\begin{eqnarray}
\mathrm{Var}\big(n_0(t)\big) = 
N_{\rm s} e^{-2 t} (1- e^{-2 t}).
\end{eqnarray}

\section{Strong uncooperative binding}
\label{appendix:sub}
\noindent
All evaluations in the main text assume $\alpha \neq 0$, since the
completely uncooperative case ($\alpha =0$) would not allow for the
formation of $s_2$ strands.  Setting $\alpha =0$ however can be used
to explore the short time behavior when $\alpha \to 0$. This is the
case of highly uncooperative binding where although rare, the
formation of a fully bound $s_2$ strand is still possible.  Setting
$\alpha = 0$ in Equations~\ref{eqn:MassActionSol}b-c and \ref{eqn:poft},
so that $e^{-\alpha t } \to 1$ for all times, we find
\begin{eqnarray}
\langle n_1 (t) \rangle &=& N_{\rm s} (1-e^{-2t}), \\
\langle n_2 (t) \rangle &=& 0 \\
\label{p0}
\langle p(t) \rangle &=& \frac 1 2 (1 - e^{-2t}).
\end{eqnarray}
Upon setting $\alpha=0$ in Equation~\ref{eqn:varm} we also find
\begin{eqnarray}
\label{varp0}
\mathrm{Var}\big(p(t)\big) &=& \frac 1 {4 N_s^2} \mathrm{Var}\big(m(t)\big) \nonumber \\
&=& \frac 1 {4 N_{\rm s}} e^{-2t} (1 -e^{-2t}) \nonumber \\
&=& \frac 1 {4 N_{\rm s}} 2 \langle p(t) \rangle (1-2 \langle p
(t) \rangle).
\end{eqnarray}
\noindent
Note that Equation~\ref{p0} yields $\langle p(t) \rangle < 0.5$
for all times, implying that for $\alpha = 0$ the reaction cannot be
completed, as expected since fully bound strands cannot emerge.
Equation~\ref{varp0} also reveals that $\mathrm{Var}\big(p(t)\big)$ is
symmetric about $ \langle p \rangle = 1/4$ and its maximum is attained
at $\mathrm{Var}\big(p(t)\big) = (16 N_{\rm s})^{-1}$.  The above
results still apply in the $\alpha \to 0$ limit, albeit for $\alpha t
\ll 1$ where $e^{-\alpha t} \to 1$.  For example
$\mathrm{Var}\big(p(t)\big)$ follows Equation~\ref{varp0} in
Figure~\ref{fgr:MeqVarCombined}b up to $\langle p(t) \rangle \sim
0.5$.  At longer times, since $\alpha$ is small but not zero, the
binding will proceed, and $s_2$ strands will emerge.  We can thus
reevaluate Equations~\ref{eqn:MassActionSol}b-c and \ref{eqn:poft},
for $\alpha \to 0$ but at long times where $e^{-2 t}
\to 0$ and $e^{-\alpha t} \neq 0$ so that
 \begin{eqnarray}
\langle n_1 (t) \rangle &=& N_{\rm s} e^{- \alpha t}, \\
\langle n_2 (t) \rangle &=& N_{\rm s} (1 - e^{- \alpha t}), \\
\langle p(t) \rangle &=&1 -  \frac 1 2 {e^{- \alpha t}}.
\label{p02}
\end{eqnarray}
\\
Finally, in the $e^{-2t} \to 0$ limit, Equation~\ref{eqn:varm} becomes
\begin{eqnarray}
\label{varp02}
\mathrm{Var}\big(p(t)\big) &=& \frac 1 {4 N_s^2} \mathrm{Var}\big(m(t)\big) \nonumber \\
&=& \frac 1 {4 N_{\rm s}} e^{-\alpha t} (1 - e^{-\alpha t}) \nonumber \\
&=& \frac 1 {4 N_{\rm s}} 2 \langle p(t) \rangle (1-2 \langle p (t) \rangle).
\end{eqnarray}
The results in Equations~\ref{p02} and \ref{varp02}
indicate that as $t \to \infty$, $\langle p(t) \rangle > 0.5$ and $\langle p(t) \rangle \to 1$.  
Furthermore, we observe that
the shape of $\mathrm{Var}\big(p(t)\big)$ in Equation~\ref{varp02} is
the same as in Equation~\ref{varp0} as also emerges from the bimodal
form in Figure~\ref{fgr:MeqVarCombined}b.  Finally, we note that for
$t \to \infty$, $\langle n_1 (t) \rangle \to 0$, even as $\alpha \to
0$ since eventually all strands will be fully bound and $\langle n_2
(t) \rangle \to N_{\rm s}$.

\bibliography{MyCollection}
\end{document}